# Smart sensors network for accurate indirect heat accounting in apartment buildings

Y. Stauffer[a], F. Saba[b,*], R.E. Carrillo[a], M. Boegli[a], A. Malengo[b], A. Hutter[a]

[a] CSEM, Centre Suisse d'Electronique et de Microtechnique, Rue Jaquet-Droz 1, 2002 Neuchâtel, Switzerland

[b] INRIM, Istituto Nazionale di Ricerca Metrologica, Strada delle Cacce 91, 10135 Torino, Italy

[*] Corresponding author.

*E-mail address*: yves.stauffer@csem.ch (Y. Stauffer), f.saba@inrim.it (F. Saba), rafael.carrillo@csem.ch (R.E. Carrillo), max.boegli@csem.ch (M. Boegli), a.malengo@inrim.it (A. Malengo), andreas.hutter@csem.ch (A. Hutter).

## Abstract

A new method for accurate indirect heat accounting in apartment buildings has been recently developed by the Centre Suisse d'Electronique et de Microtechnique (CSEM). It is based on a data driven approach aimed to the smart networking of any type of indirect heat allocation devices, which can provide, for each heat delivery point of an apartment building, measurements or estimations of the temperature difference between the heat transfer fluid and the indoor environment. The analysis of the data gathered from the devices installed on the heating bodies, together with the measurements of the overall building heat consumption provided by direct heat metering, allows the evaluation of the characteristic thermal model parameters of heating bodies at actual installation and working conditions. Thus overcoming the negative impact on accuracy of conventional indirect heat accounting due to off-design operation, in which these measurement systems normally operate. The method has been tested on conventional heat cost allocators (HCA), and on innovative smart radiator thermostatic valves developed by CSEM. The evaluations were carried out at the centralized heating system mock-up of the Istituto Nazionale di Ricerca Metrologica (INRIM), and also in a real building in Neuchatel, Switzerland. The method has proven to be an effective tool to improve the accuracy of indirect heat metering systems; compared to conventional HCA systems, the error on the individual heating bill is reduced by 20% to 50%.

## Nomenclature

| | |
|---|---|
| CSEM | Centre Suisse d'Electronique et de Microtechnique |
| DHM | Direct Heat Meters |
| EDD | European Directive on Energy Efficiency |
| HCA | Heat Cost Allocator |
| INRIM | Istituto Nazionale di Ricerca Metrologica |
| ITC | Insertion Time Counters |
| MAPE | Mean Absolute Percentage Error |
| MID | Measuring Instrument Directive |
| RED | Remote Ethernet Device |
| RLS | Regularized Least Squares |
| RSP | Room temperature Set-Point |
| PSP | Valve Position Set-Point |
| SMINTEBI | Smart Individual Tenant Billing system |
| STV | Smart Thermostatic Valves |

## 1. Introduction

Nowadays, the European building stock constitutes an energy system with a significant environmental impact, accounting for about 40% of primary energy consumption and 36% of $CO_2$ emissions [1]. The European Directive on Energy Efficiency 2012/27/EU (EED) [2] and its subsequent recast 2018/844/EU [3] target energy savings in buildings with a number of actions, among which the accurate estimation of individual thermal energy consumptions, and consequently, the fair heat cost allocation among the residents of apartment buildings with centralized heating systems. Accurate heat metering and heat allocation are, undoubtedly, important driving forces towards energy saving, energy efficiency and the reduction of pollutant emissions of buildings.

Direct heat metering is certainly the preferred and most accurate heat accounting method, providing the direct measurement of the thermal energy transferred by a fluid flow passing through a generic heat exchanger. In particular, Direct Heat Meters (DHM) measure the thermal energy transferred by a fluid flow, by the time integration of the product between the mass flow rate of the heat transfer fluid and its specific enthalpy difference between the inlet and outlet flow sections of the heat exchanger. On the technical point of view, DHMs are regulated by the OIML R 75 [4] and the European Standard EN 1434 [5]. Moreover, in order to ensure customer protection through fair measurements and to eliminate trade barriers among member states, the 2004/22/CE Measuring Instrument Directive (MID) [6], which harmonizes the measuring instrument legislation in Europe, regulates and fixes the rules for the approval and initial verification of DHMs (Annex MI-004).

However, in buildings with vertical risers hot water distribution networks, the use of DHMs for the correct allocation of heat consumptions among the users can be practically and economically unfeasible because of installation constraints and high costs, due to necessity of installing a DHM for each heating body. This is true especially for retrofitting cases, where the labour cost in addition to the hardware cost is prohibitive. In such cases, the EED allows using indirect heat accounting systems, which carry out an estimation of individual thermal energy consumptions, based on the approximated evaluation of the temperature difference between the heat transfer fluid and the indoor environment at the different heat delivery points. The two main typologies of indirect heat accounting systems available on the market are the heat cost allocators (HCA) and the insertion time counters (ITC) [7]. Unlike DHMs, they are not regulated by the MID, but are mandatory in order to improve energy efficiency and promote energy saving. Of course, in order to ensure consumer protection, indirect heat accounting devices must be compliant to proper regulations and technical standards, like the EN 834 [8] for heat cost allocators and the UNI 11388 [9] for ITCs. HCAs and ITCs are based on the indirect evaluation of the heat output of radiators and convectors, by means of their nominal thermal power, determined according to the European Standard EN 442 [10], and by the estimation of the temperature difference between the heat transfer fluid and the room environment. In particular, HCAs provide an evaluation of such a temperature difference by correcting the measured temperature difference between radiator surface and room environment using proper coupling parameters, which depend on the type of radiator and the type of HCA (i.e. HCA with embedded or separate room temperature sensor). On the other hand, for ITCs, the mean fluid temperature is evaluated from the main forward and return temperature of the heating system.

According to the European Standard EN 442, the indirect evaluation of the thermal power $q_{rad}$ emitted in steady-state conditions by radiators and convectors is obtained by the following model:

$$q_{rad} = q_{N,50} \left( \frac{T_m - T_a}{50} \right)^n \qquad (1)$$

where $T_m$ denotes the mean fluid temperature evaluated by the temperatures measured at the inlet and outlet flow section of the radiator, $T_a$ the air temperature near the surface of the radiator, whereas $q_{N,50}$ and $n$ are the characteristic thermal model parameters of the radiator. In particular, $q_{N,50}$ represents the radiator nominal thermal power exchanged when $T_m - T_a$ = 50 °C, whereas the exponent $n$ typically ranges from 1.1 to 1.4 depending on the ratio between the thermal output exchanged by radiation and convection. The characteristic radiator parameters are obtained by regression analysis of the experimental data of $q_{rad}$ and $(T_m - T_a)$, which are measured in a thermal chamber at specified installation and heat transfer conditions.

However, in real life applications, due to the very common off-design installation and operation of heating bodies with respect to the reference laboratory conditions defined by the EN 442 Standard, the validity of nominal thermal model parameters of radiators can be significantly compromised, so that the accuracy of typical indirect heat accounting methods decreases [12]. Relative deviations higher than 20% have been observed for indirect estimations of radiator thermal power at typical operating conditions. This deviation is due to several installation effects, like different inlet-outlet piping connection layouts, installation position with respect to the wall and the floor, niche confinement, presence of grid, shelves or other obstructions close to radiator surfaces (curtains covering), and variation of thermo-fluid-dynamic conditions (inlet flow rate and temperature) [12]. Other remarkable installation effects that can negatively affect indirect heat accounting methods are the presence of air and impurities inside the radiators, the variation of surface emissivity (painting of radiator surface), and the effect of atmospheric pressure variation as a function of altitude [12]. In addition, it is quite common, particularly for outdated radiator types, that characteristic thermal model parameters are not available for the installer of indirect heat accounting systems. In this case, the parameters have to be obtained by comparison with similar characterized radiators, or by means of empirical methods like the one proposed by the Italian Standard UNI 10200 [15], an increased uncertainty of thermal output estimation, of more than 10%, should be considered [13].

In order to overcome the aforementioned issues related to indirect heat accounting systems and to promote energy efficiency in buildings [16], a new method, able to estimate the characteristic thermal model parameters of radiators and convectors during the real operation of the heating system, has been developed at the Centre Suisse d'Electronique et de Microtechnique (CSEM). The method is based on the smart networking of indirect heat accounting devices, which are able to provide measurements or proper approximations of the temperature difference between the heat transfer fluid and the indoor environment at each heat delivery point of a residential building. In addition, the method also requires the overall thermal energy measurements provided by a central DHM installed on the heater supply line, which are acquired for different working conditions of the heating system, and used for the estimation of the actual characteristic thermal model parameters of the heating bodies. Such parameters, according to the model used for the indirect evaluation of radiators or convectors thermal output, can be defined as the thermal powers exchanged by the heating bodies at a specified value of the temperature difference sensed by the indirect heat accounting devices. The parameters' estimation is posed as a linear inverse problem and solved by means of a Tikhonov regularization (Regularized Least Squares) algorithm [17]. The method has been applied both to a typical HCA system, and to a novel heat accounting and temperature control system, the Smart Individual Tenant Billing system (SMINTEBI), based on the use of radiator Smart Thermostatic Valves (STV). Tests have been carried out at the centralized heating system mock-up of the Istituto Nazionale di Ricerca Metrologica (INRIM), where the sensitivity of the method to the main uncertainty sources has been analysed, and the accuracy of heat accounting has been evaluated by comparison against traceable thermal energy measurements provided by direct heat meters reference standards. Finally, in order to assess the effectiveness of the method in real conditions and its robustness to user and weather disturbances, a preliminary pilot test of the SMINTEBI system was carried out in a building in Neuchatel, Switzerland.

In Section 2, the novel heat accounting method is detailed, whereas in Section 3, the design of tests, carried out at the INRIM centralized heating system mock-up for the validation of the method, is described. The results of the experimental validation carried out at INRIM, and the results of the pilot test of the SMINTEBI system performed on a real operational site in Neuchatel are reported and discussed in Section 4.

## 2. Methodology

In this section, we detail the proposed indirect heat accounting estimation method. We start by introducing some preliminaries in indirect heat accounting devices and follow by the description of the proposed approach and its application to two particular smart sensor networks. The overall approach is depicted in Fig. 1.

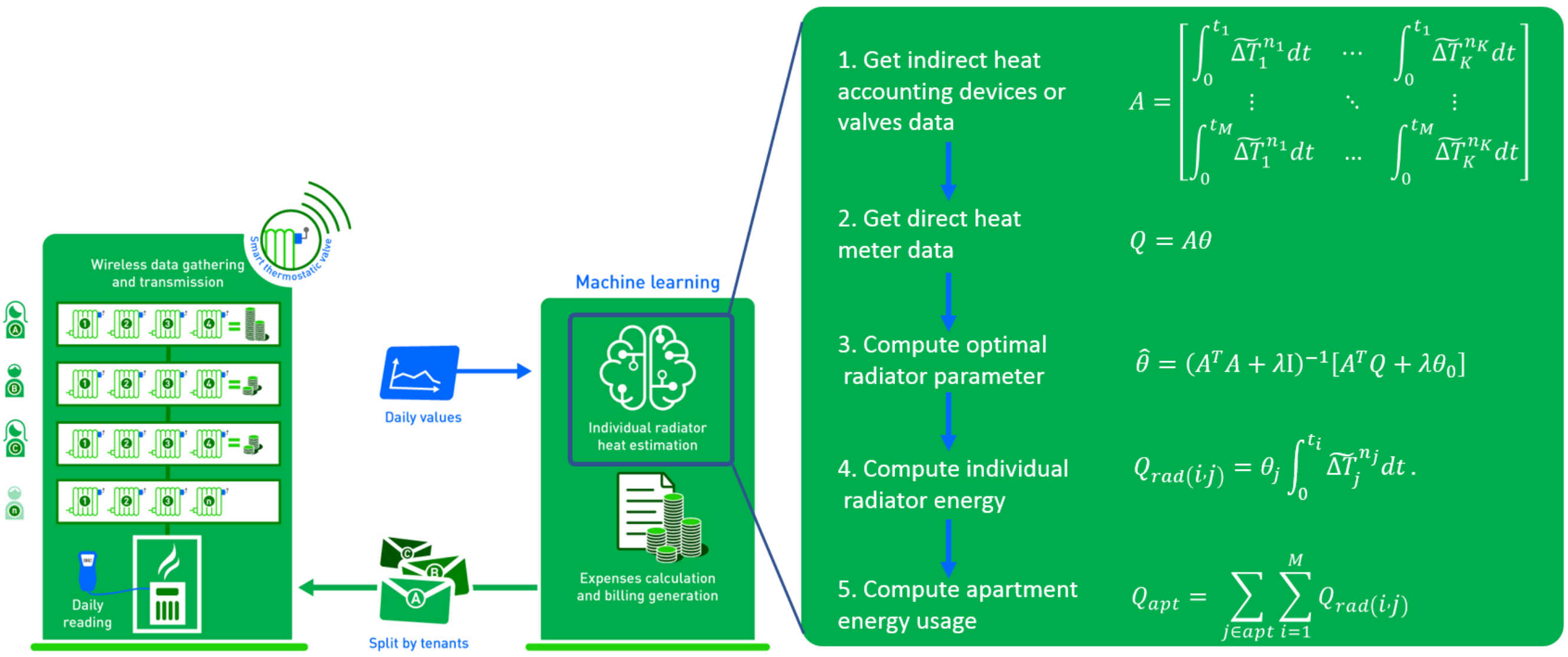

**Fig. 1.** Schematic representation of the overall approach of the proposed indirect heat accounting method.

### *2.1. Indirect heat accounting devices*

Indirect heat accounting systems provide an estimation of individual thermal energy consumptions in terms of number of “allocation units”, allowing the sharing of the heating costs among the tenants, as a fraction of the whole thermal energy consumption of the building (usually measured by gas meters or DHMs).

The two main typologies of indirect heat accounting systems available on the market are described in this section: the heat cost allocators (HCA), and the insertion time counters (ITC). Both these devices provide an indirect estimation of the thermal energy exchanged by water radiators, whose nominal thermal output characteristics are determined according to the European Standard EN 442 [10].

Since the “allocation units” provided by indirect heat accounting devices cannot be considered as measurements of the actual amount of individual heat consumption expressed in energy units, they can be used, exclusively, to assess the sharing of the variable part of the thermal energy bill among the tenants of apartment buildings. The split of the total heating cost in a fixed part, related to the heat losses through the circuit and the heater, and a variable or voluntary part, related to the actual individual heat consumptions, is usually based on the characteristic energy performance and the typical occupancy behaviour of the building. Once such fixed and variable fractions have been defined, the fixed part of the overall thermal energy bill of the building is usually divided among the users according to the size of their apartments, while the variable part is divided among the tenants according to the corresponding number of "allocation units" provided by the indirect heat accounting devices.

HCAs provide an indirect estimation of radiators and convectors heat consumption, which is obtained from i) their nominal thermal power, determined accordingly to the EN 442 standard, ii) proper empirical coefficients that take into account the HCA-radiator coupling, and iii) the time integration of the measurement of the temperature difference between a specified point on the radiator surface and the indoor ambient air in proximity of the radiator. HCAs are regulated by the European Standard EN 834 [8], which fixes the requirements and the testing procedures for such devices. The EN 834 standard describes three measurement methods for HCAs: single-sensor, two-sensors and multiple sensors. HCAs are designed to provide a totalized number of allocation units, which is proportional to the thermal energy exchanged by water radiators, according to an approximation of the EN 442 model given by equation (1). In fact, the measured temperature difference between the radiator surface and the surrounding air is intended to approximate the difference between the mean temperature of the heat transfer fluid inside the radiator and the room temperature. Of course, different correction factors are needed, in order to take into account the effect of the thermal coupling of HCA temperature sensors on different types of radiators, at fixed attachment position of the sensors and heat transfer conditions.

The measurement model implemented by HCAs for the evaluation of the totalized number of allocation units $R$ can be expressed as follows [7]:

$$R = K_Q K_C K_T \int \left(\frac{T_s - T_{ar}}{60}\right)^n dt \tag{2}$$

where $K_Q$ is the rating factor for the nominal thermal output of the radiator, $K_C$ the rating factor for the thermal coupling of the temperature sensors, $K_T$ the rating factor for lower room design temperatures, $T_s$ the radiator surface temperature measurement, $T_{ar}$ the room air temperature measurement, $n$ the radiator characteristic exponent (usually considered constant and equal to 1.3), and $t$ the time. The rating factor $K_Q$ coincides with the radiator nominal thermal power $q_{N,60}$ exchanged at a temperature difference of 60 °C between the heat transfer fluid and the surrounding air. It can be obtained from the EN 442 radiator nominal thermal power $q_{N,50}$ by the relation $K_Q = q_{N,60} = q_{N,50}(60/50)^n$. The rating factor $K_C$ is the correction factor that takes into account the different thermal coupling of the HCA temperature sensors for different types of heating surfaces, whereas the rating factor $K_T$ takes into account the change of the thermal output of water radiators when single sensor HCAs are used at design indoor temperatures less than the basic reference air temperature.

The accuracy of HCAs is significantly affected by their correct parameterisation in terms of the rating factors $K_Q$ and $K_C$. In those installation cases in which the nominal thermal model parameters of the radiator are not available and the radiator type does not clearly belong to any defined category (outdated radiators), the installer can approximatively determine the radiator nominal thermal power by means of a proper dimensional method [15] and parametrize HCAs following empirical considerations on the similarity with other defined radiator types, using proper rating factors databases. Such common installation situations, in addition to the application of HCAs on radiators characterized by operating conditions different from the ones of the reference thermal chamber in which they were characterized, may lead to considerable measurement errors.

Concerning ITCs, they are used in heating systems controlled by zone valves or by ON-OFF valves installed on each water radiator and are regulated, in Italy, by the UNI TR 11388 Technical Report [9]. ITCs provide the indirect estimation of the thermal energy exchanged by water radiators by means of the measurement of insertion time of radiators, that is the time during which the zone valves or the radiator valves remain open, and the characteristic thermal model parameters of water radiators, $q_{N,50}$ (or $q_{N,60}$) and $n$, determined accordingly to the EN 442 standard. In particular, ITCs estimates the thermal output of water radiators during both the opening time of the corresponding valves and the time in which they are kept closed just after a heating period.

### 2.2. Improved method for the realization of smart networks of indirect heat accounting devices

The indirect evaluation of radiators and convectors heat consumption in real life operating conditions, by using the nominal thermal outputs obtained in reference laboratory conditions according to the EN 442 Standard, can lead to measurement errors depending on the particular radiator installation and working conditions. For this reason, the typical relative uncertainties associated to the estimations of heat consumption provided by conventional indirect heat accounting systems are around 10% for HCAs and between 13% and 20% for ITCs [18].

The method developed by the Centre Suisse d'Electronique et de Microtechnique (CSEM) for indirect heat accounting in apartment buildings ensures the on-site identification of the characteristic thermal model parameters of each heating body, providing the evaluation of their heat consumption under actual installation and working conditions. This is possible by solving the energy balance applied to the building thermal-hydraulic distribution, by means of:

- measurements of the overall thermal energy exchanged by the heat transfer fluid in the heating system, as provided by a DHM installed on the main flow line of the thermal-hydraulic distribution,
- measurements or proper approximations of the temperature difference between the heat transfer fluid and the room environment at each heating body of the apartment building, as provided by indirect heat accounting devices installed on the heating bodies.

Heat losses through the pipework and transient contributions associated to the thermal inertia of the system, due to the complexity of a deterministic approach in real operational sites, are treated as uncertainty sources, and their influence on the accuracy of the method has been evaluated experimentally as described hereinafter in Section 4.1.1.1.

To properly estimate the thermal model parameters of all the heating bodies, the energy balance is solved as a linear inverse problem, using a set of measurement data representing different working conditions of the heating system. The radiator thermal model parameters, according to the model described in Section 2.2.1 for the indirect evaluation of radiators or convectors thermal output, are defined as the mean thermal powers exchanged by the heating bodies at a specified value of the temperature difference between radiator and room environment, under the actual installation and operating conditions in which the radiator is used. The parameters estimates are used to compute the individual heat consumptions, and can be updated or re-calibrated on a regular basis, e.g. every heating season or every month, in order to take into account either changes of installation and working conditions, or possible drift effects of the measurement systems.

The aforementioned approach allows establishing a smart sensor network of the indirect heat accounting devices installed at radiator level, combining their readings according to the energy balance of the thermal hydraulic distribution, in order to provide the on-site calibration of the characteristic thermal model parameters of heating bodies.

In the following sections, the novel improved heat accounting method is described from the mathematical point of view, and its possible applications to typical HCA systems and to a novel indirect heat accounting system based on smart radiator thermostatic valves, are presented.

### 2.2.1. Calculation method

Suppose there are $K$ radiators connected to the same flow line of the thermal-hydraulic distribution and that the overall thermal energy at this flow line level is measured for a certain period, for example daily or every few hours, and this measurement is repeated for $M$ periods, preferably under different working conditions of the heating system, e.g. $M$ days during the heating season. Let $Q_{rad(i,j)}$ denotes the thermal energy exchanged by the $j$-th radiator ($j = 1, \dots, K$) in the $i$-th period ($i = 1, \dots, M$), which can be estimated by an indirect heat accounting device as:

$$Q_{rad(i,j)} = \theta_j \int_0^{t_i} \widetilde{\Delta T}_j^{\,n_j} dt \tag{3}$$

where $\theta_j$ is the characteristic thermal model parameter of the $j$-th radiator, $\widetilde{\Delta T}_j$ is the temperature difference sensed by the indirect heat accounting device (between the heat transfer fluid or radiator surface and the room environment), normalized with respect to a base temperature difference (equal to 50 °C according to the EN 442 model of equation (1), or 60 °C according to the HCA calculation model of equation (2)), $n_j$ is the characteristic radiator exponent (considered as a constant parameter usually equal to 1.3), and $t_i$ is the duration of the $i$-th integration period.

Let $\boldsymbol{Q}$ be an $M$-dimensional vector containing the values of the overall thermal energy measurements provided by the main DHM, for the $M$ energy samplings (e.g. daily consumptions). Defining $\boldsymbol{\delta Q}$ as an $M$-dimensional vector accounting for heat losses through the pipework, thermal inertia of the system, and DHM measurement uncertainty, the following linear model, which describes the energy balance applied to the thermal-hydraulic distribution, can be posed:

$$\boldsymbol{Q} = \boldsymbol{A\theta} + \boldsymbol{\delta Q} \tag{4}$$

where $\boldsymbol{\theta}$ is a $K$-dimensional vector of the $\theta_j$ parameters, and the matrix $\boldsymbol{A}$ is an a $M \times K$ matrix defined as:

$$\boldsymbol{A} = \begin{bmatrix} \int_0^{t_1} \widetilde{\Delta T}_1^{\,n_1} dt & \cdots & \int_0^{t_1} \widetilde{\Delta T}_K^{\,n_K} dt \\ \vdots & \ddots & \vdots \\ \int_0^{t_M} \widetilde{\Delta T}_1^{\,n_1} dt & \dots & \int_0^{t_M} \widetilde{\Delta T}_K^{\,n_K} dt \end{bmatrix} \tag{5}$$

To estimate the parameters $\theta_j$ for all the $K$ radiators, we need to estimate $\boldsymbol{\theta}$ from the linear inverse problem posed in equation (4).

Note that the inverse problem in equation (4) can be ill-posed if $M < K$ and/or if the rows or columns of $\boldsymbol{A}$ are correlated, which is the normal case if the measurements are performed in normal operating conditions. Thus, prior information about the problem needs to be used to be able to solve the inverse problem. In order to account for prior information on the parameters $\theta_j$, the vector $\boldsymbol{\theta_0}$ is considered as prior information. The prior information might be the manufacturer's parameters, parameters from similar radiators or parameters computed in a previous season. The $\theta_j$ parameters are determined by solving the following regularized least squares (RLS) problem [17]:

$$\min_{\boldsymbol{\theta}}(\|\boldsymbol{Q}-\boldsymbol{A\theta}\|_2^2+\lambda\|\boldsymbol{\theta}-\boldsymbol{\theta_0}\|_2^2) \tag{6}$$

where $\|\cdot\|_2^2$ represents the squared L2 norm for vectors. The first term in equation (6) measures the discrepancy between the linear model and the measured thermal energy, whereas the second term measures the discrepancy between the estimated parameters $\boldsymbol{\theta}$ and the prior information $\boldsymbol{\theta_0}$. The parameter $\lambda$ is a regularization parameter that balances the influence of the prior information and the measured thermal energy data. The value of $\lambda$ is adapted to the particular situation, e.g. large values of $\lambda$ give more trust to the prior information $\boldsymbol{\theta_0}$ or small values of $\lambda$ give more trust to thermal power data. It is important to observe that the term $\boldsymbol{\delta Q}$, due to the complexity of a deterministic approach in real operational sites, is treated as a zero mean uncertainty term, and does not appear in equation (6). The influence on the accuracy of the method, due to heat losses through the pipework and transient contributions associated to the thermal inertia of the system, has been evaluated experimentally as described hereinafter in Section 4.1.1.1. The problem in equation (6) has the closed-form solution [17]:

$$\widehat{\boldsymbol{\theta}}=(\boldsymbol{A}^T\boldsymbol{A}+\lambda\mathbf{I})^{-1}[\boldsymbol{A}^T\boldsymbol{Q}+\lambda\boldsymbol{\theta_0}]. \tag{7}$$

Taking a probabilistic point of view of the problem [17], the covariance matrix of the estimate $\widehat{\boldsymbol{\theta}}$ is:

$$\boldsymbol{C}_{\widehat{\boldsymbol{\theta}}}=(\boldsymbol{A}^T\boldsymbol{A}+\lambda\mathbf{I})^{-1}. \tag{8}$$

The covariance matrix (especially its diagonal) quantifies the variances of the parameters estimates.

The estimation method enables to update the parameters $\widehat{\boldsymbol{\theta}}$ on a regular time basis, e.g. using weekly or monthly data, or to estimate them by using the whole data batch from the heating season, in order to evaluate the radiator heat consumption either in specific periods, or in the whole heating season.

To evaluate the regularization parameter $\lambda$, we adopted the L-curve method [20]. The method consists in tracing the L-curve by plotting in double logarithmic scale, for different values of $\lambda$, the norm of the deviation between parameters estimates and their prior knowledge $\|\boldsymbol{\theta}-\boldsymbol{\theta_0}\|_2$, versus the corresponding norm of residuals $\|\boldsymbol{Q}-\boldsymbol{A\theta}\|_2$: the regularization parameter corresponding to the point at the maximum curvature of the L-curve approximates the optimal regularization parameter.

### *2.2.2. Application to HCA systems*

An important advantage of this novel heat accounting method is that it is applicable to traditional indirect heat accounting devices. Concerning HCA systems, the method allows converting the set of devices in a smart sensors network, where each HCA reading is corrected or calibrated accordingly to the actual installation and heat transfer conditions of the corresponding radiator. The characteristic thermal model parameters of heating bodies are obtained by the solution of the RLS problem described by equation (6), which represents the best fit with respect to the energy balance of the thermal-hydraulic distribution at different working conditions.

The procedure for estimating the characteristic thermal model parameters of radiators in a system equipped with HCAs can be described as follows:

a. HCAs are parameterized setting the rating factors $K_Q$ = $K_C$ = $K_T$ = 1, so that their outputs are functions exclusively of the temperature difference between the radiator surface and the indoor environment (the characteristic radiator exponent $n$ is considered constant), as can be observed from equation (2).
b. The readings of HCAs and the energy output of the DHM installed on the main flow line of the thermal-hydraulic distribution are acquired regularly over the heating season, i.e. daily, in order to gather a batch of data, which is representative of the variety of working conditions of the centralized heating system. The measurement data from HCAs and DHM are used to form, respectively, the matrix $\boldsymbol{A}$ and the vector $\boldsymbol{Q}$ shown in equations (4) and (5). In particular, equation (3), which describes the thermal energy $Q_{rad(i,j)}$ exchanged by the *j*-th radiator in the *i*-th integration period, takes the following form:

$$Q_{rad(i,j)} = \theta_j \hat{R}_{i,j} \tag{9}$$

where $\hat{R}_{i,j}$ is the number of allocation units totalized by the HCA installed on the *j*-th radiator during the *i*-th integration period, configured with rating factors $K_Q$ = $K_C$ = $K_T$ = 1.
c. The characteristic parameters $\theta_j$ of each radiator are estimated by solving the RLS problem described by equation (6), where the vector $\boldsymbol{\theta_0}$ contains the prior knowledge about the radiators parameters, which can be considered equal to the product of the rating factors $K_Q$, $K_C$, and $K_T$, determined accordingly to the nominal configuration of HCAs (based on the EN 834 Standard and radiator manufacturer's data).

The accuracy of the method applied to HCAs has been tested at the INRIM centralized heating system mock-up, where the influence of the uncertainty sources associated to the heat losses through the pipework and the thermal inertia of the system has been investigated. In addition, the sensitivity of the method to different values of $\boldsymbol{\theta_0}$ has been analysed to take into account the effect of possible errors on the nominal configuration parameters of HCAs. The results will be shown in Section 4.1.1.1.

### 2.2.3. *Application to the SMINTEBI heat accounting system*

Besides the application to typical HCA systems, the improved heat accounting method has been also implemented on a novel system of radiator Smart Thermostatic Valves (STV) to form the Smart Individual Tenant Billing system (SMINTEBI), which allows the combined heat accounting and room control temperature in residential buildings.

The SMINTEBI heat accounting realises on the following elements:

- Sensing at radiator level by the SmartDrive-MX STV from HORA (Fig. 2a). These valves measure the room temperature (a temperature sensor is located on the PCB extremity that is thermally insulated from the radiator), radiator inlet fluid temperature (a temperature sensor is located on the PCB located close to the radiator in the vicinity of the metallic valve winding) and valve opening (i.e. position in %, the maximal stroke is 5.5 mm with a resolution of 33 μm). The valves are configured to provide these measurements every 5 minutes.
- Sensing at the heater level by a DHM installed on the main flow line of the thermal-hydraulic distribution, in order to measure the overall thermal energy exchanged by the heat transfer fluid.
- STVs and DHM logging and control, by means of a standard PC equipped with a USB to EnOcean dongle (Fig. 2b), which is used to control the valves (i.e. regulate the opening as required as a function of the desired room temperature) and retrieve the measured data before sending them to a database located at CSEM facilities through a VPN tunnel created by a Remote Ethernet Device (RED).

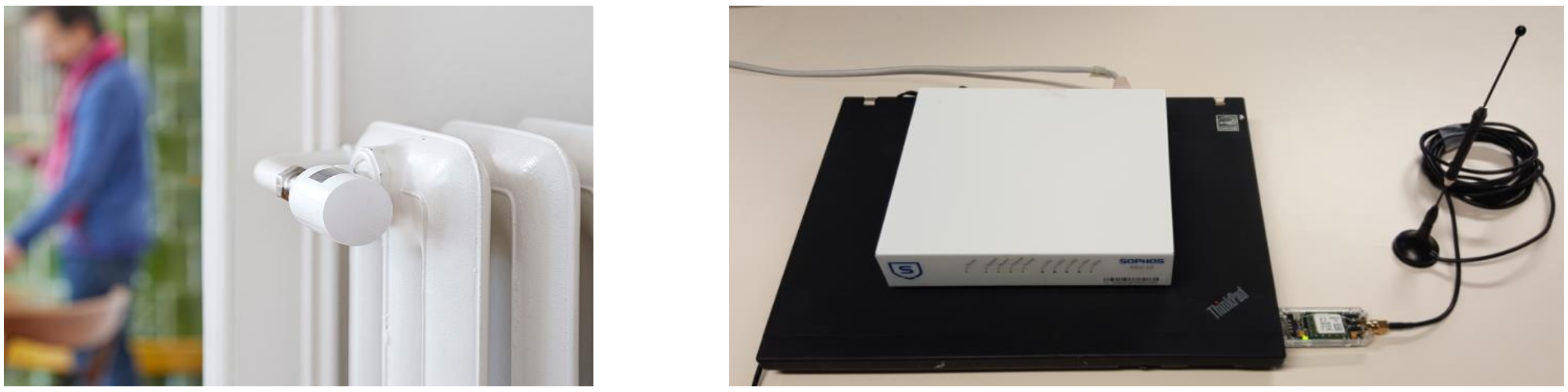

**(a)** **(b)**

**Fig. 2.** (a) SmartDrive-MX STV from HORA; (b) logging PC with EnOcean dongle and RED.

The measurement data from STVs and DHM are used to form, respectively, the matrix $\boldsymbol{A}$ and the vector $\boldsymbol{Q}$ shown in equations (4) and (5). In particular, equation (3), which describes the thermal energy $Q_{rad(i,j)}$ exchanged by the $j$-th radiator in the $i$-th integration period, takes the following form:

$$Q_{rad(i,j)} = \theta_j \int_0^{t_i} \left( \frac{T_{in,j} - T_{av,j}}{50} \right)^{n_j} dt \tag{10}$$

In equation (10), the characteristic radiator exponent $n_j$ of the $j$-th radiator is considered constant (equal to 1.3), whereas $T_{in,j}$ and $T_{av,j}$ are, respectively, the fluid inlet temperature and the room temperature measured by the STV installed on the $j$-th radiator. It is important to observe that the estimated parameters $\theta_j$ have a different meaning from the EN 442 nominal radiator thermal powers $q_{N,50}$, since $T_{in,j}$ does not coincide with the mean temperature of the heat transfer fluid inside the radiator. In addition, since the model described by equation (10) does not allow to clearly discriminate between partial and full STV openings (at constant heater flow temperature, the inlet radiator temperatures are expected to be independent from STVs percentage opening), $\theta_j$ can be considered only as mean parameters, averaged over the radiators working conditions and the ON-OFF insertion times of STVs during the $i$-th integration period.

The parameters $\theta_j$ of each radiator are estimated by solving the RLS problem described by equation (6), where the elements of the initial guess vector $\boldsymbol{\theta_0}$ can be reasonably considered as equal to the nominal radiator thermal powers $q_{N,50}$, provided by manufacturers according to the EN 442 Standard.

The advantages of the SMINTEBI system are that it does not require any temperature measurement of the radiator surface, and, heat accounting and room temperature control are provided at once by the same device. However, the drawback with respect to the application of the method to HCAs is that the measurement model given by equation (10) is expected to be less accurate than the one given by equation (9), as confirmed by the experimental results shown in Section 4.

## 3. Experimental validation

The experimental validation of the improved heat accounting method has been carried out in 2019 at the INRIM centralized heating system mock-up, where the accuracy of the share of heat consumption provided by the method applied both to HCAs and STVs systems has been compared against the accuracy of the heat allocation obtained by HCAs (at nominal parameters configuration). The accuracy of the heat allocation provided by the three indirect heat accounting systems (improved heat accounting method applied to HCAs, SMINTEBI system, and HCAs) has been evaluated by comparison against the share of heat consumption obtained by traceable thermal energy reference standards (reference DHMs installed on each radiator).

A test plan of a total duration of 23 days has been properly designed to validate the new heat accounting method at the following typical operating conditions:

- heating system consisting of different types of radiators;
- different types of heater working conditions: constant *vs* variable heater forward temperature;
- different types of radiator valve controls: pattern based position *vs* temperature controlled.

### *3.1. Test facility*

The INRIM centralized heating system mock-up is a laboratory for testing and validating heat accounting systems and methods at actual operating conditions. It is a full-scale centralized heating system with 40 water radiators of different types and dimensions, connected to an automatically reconfigurable hydraulic circuit (Fig. 3).

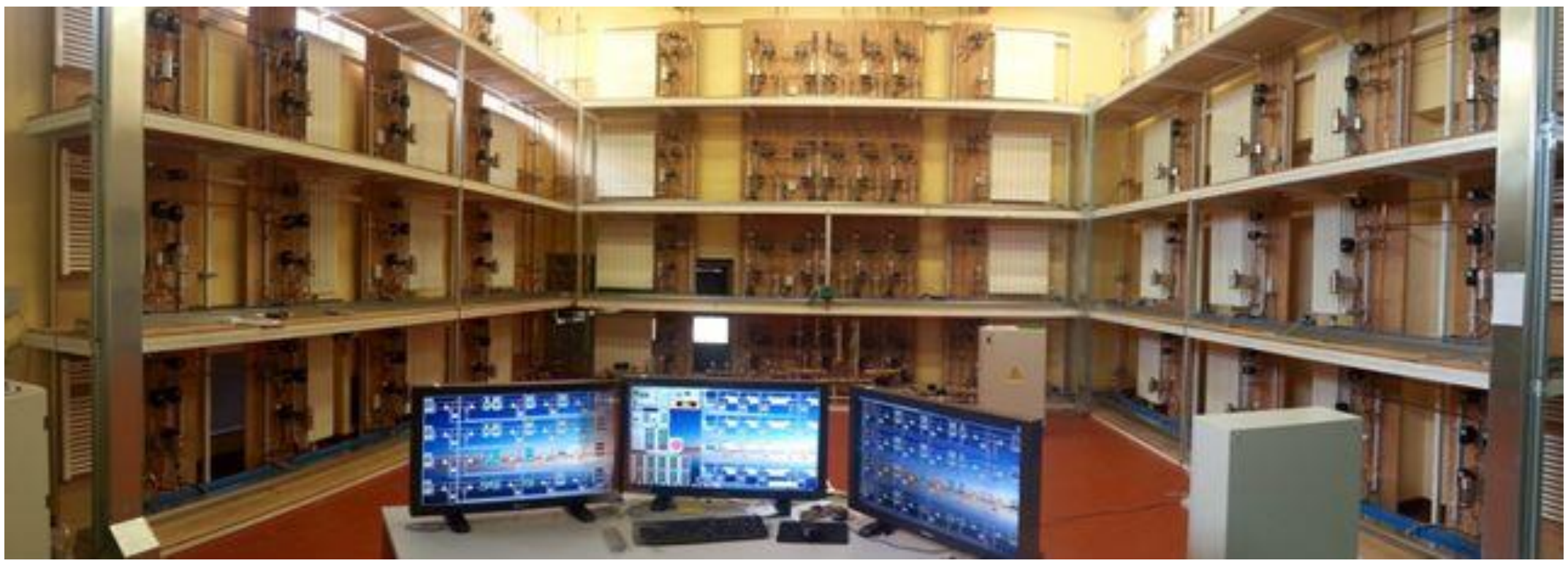

**Fig. 3.** The INRIM centralized heating system mock-up.

A natural gas condensation boiler, with thermal power modulation of the burner ranging from about 10 kW to 50 kW, and a centrifugal pump with electronic speed control are used for generating the desired water supply temperature, flow rate and pressure inside the thermal-hydraulic circuit. The adequate air exchange rate of the laboratory and the indoor temperature control is ensured by a mechanical ventilation system.

The configuration of the hydraulic circuit can be modified by means of 180 motorized on-off valves, in order to connect the water radiators by main risers vertical distribution, or horizontal loops (parallel and serial connection). In Fig. 4, the different types of radiator connected by vertical hydraulic distribution are shown: it can be observed that the circuit consists of two main groups of 20 radiators, the "north wall" and the "south wall", which are characterized by a mirrored arrangement of radiators of the same type. Each water radiator has a motorized on-off inlet valve and is equipped with a threaded connection for the installation of typical thermostatic valves. A software allows configuring the hydraulic distribution layouts and enables to control and monitor the whole operation of the central heating system.

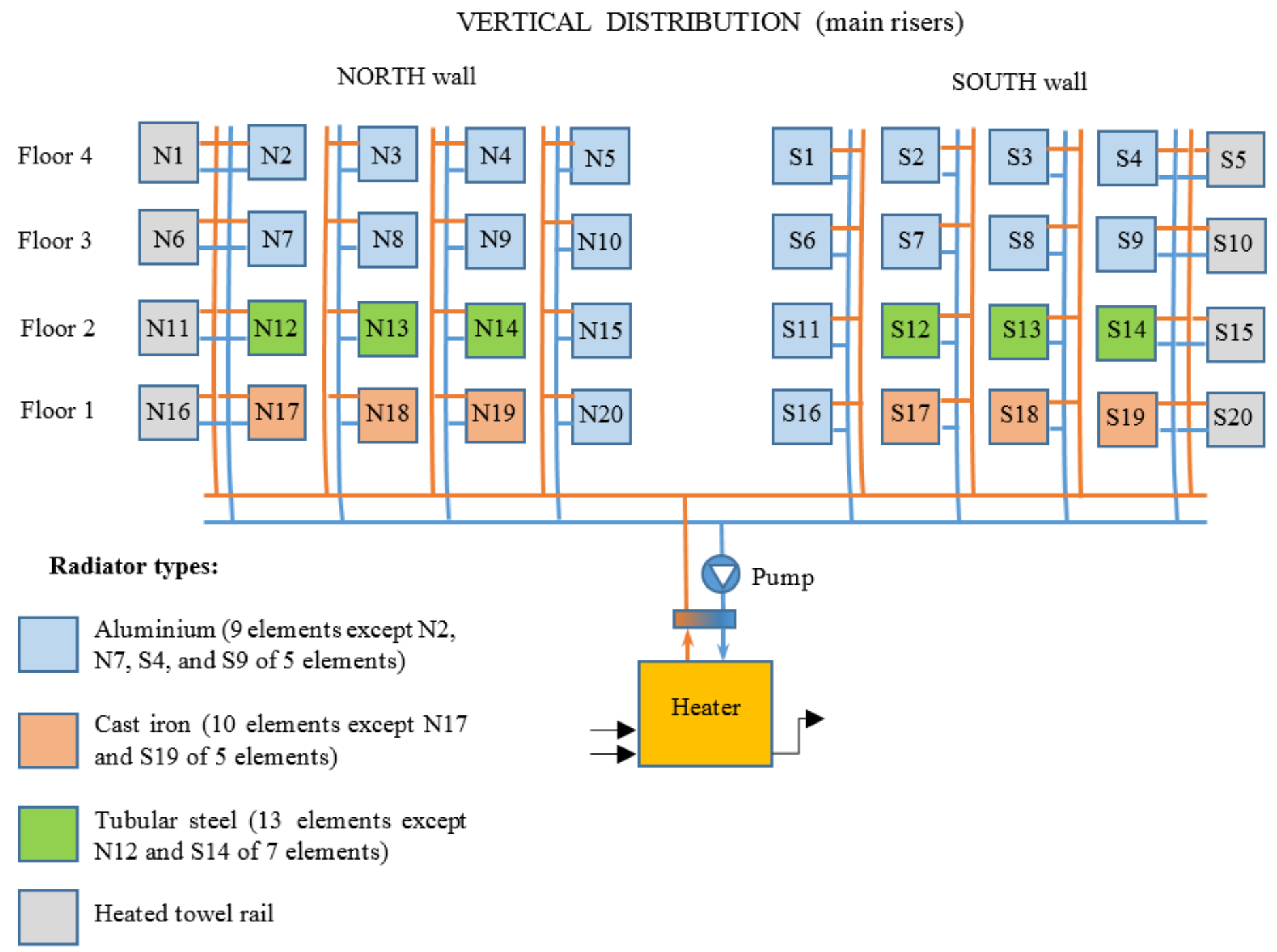

**Fig. 4.** Main risers vertical hydraulic distribution and radiators arrangement at the INRIM centralized heating system mock-up.

Size and nominal thermal characteristics of water radiators installed at the INRIM laboratory are listed in Table 1.

**Table 1.** Characteristics of radiators installed at the INRIM centralized heating system mock-up.

| Radiator type | Number of radiators in the laboratory | Number of elements of the radiator | Element dimensions (H x L x W) / mm | Nominal thermal power of radiator element at ΔT=50 °C (EN 442 [10]) / W | Radiator exponent (EN 442 [10]) |
|---|---|---|---|---|---|
| Cast aluminium sectional | 16<br>4 | 9<br>5 | 875 x 80 x 80 | 163 | 1.359 |
| 4 columns cast iron sectional | 4<br>2 | 10<br>5 | 875 x 60 x 130.5 | 142.7 | 1.3679 |
| 4 columns tubular steel | 4<br>2 | 13<br>7 | 900 x 46 x 136 | 114 | 1.28 |
| Heated towel rail | 8 | - | 713 x 535 x 30 | 395 | 1.25 |

Each water radiator is equipped with its reference DHM for the measurement of the thermal energy exchanged by the heat transfer fluid flowing through the radiator. Each radiator DHM consists of the following sub-assemblies:

- A DN8 electromagnetic flow meter with a measurement range from (2.5 to 500) L $h^{-1}$,
- A pair of PT100 thermometers for inlet and outlet temperature measurement of the fluid through the radiator (temperature sensors directly immersed in the fluid without pockets),
- A thermal power and thermal energy calculator (Radiator Calculation Unit), implementing the reference formulation of the equation of state for the evaluation of water thermodynamic properties [22].

Measurements of radiators thermal power, thermal energy, flow rate, inlet and outlet temperatures are provided every 15 s at least.

Traceability of thermal energy measurements is ensured by periodical calibration of flow meters and temperature sensor pairs of radiator DHMs against the corresponding INRIM flow rate and temperature national standards.

Each radiator of the heating system is also equipped with its HCA, installed according to the EN 834 Standard [8]. The acquisition of HCA readings is carried out by means of a radio receiver, at a maximum frequency of one reading every 3 hours.

The validation of indirect heat accounting systems and methods is carried out by comparison against the thermal energy reference standards at steady-state conditions or considering an overall test duration, which includes both the transient phases of heating and cooling of radiator surfaces.

It is worth observing that the radiators of the INRIM central heating system mock-up are characterized by different installation and heat transfer conditions with respect to the ones of the EN 442 reference thermal chamber, in which the radiator nominal parameters are determined. In particular, enhanced convective heat transfer conditions are expected for radiators closer to the air intakes of the ventilation system. Moreover, the presence of radiators of the same type, but with a different number of elements, allows taking into account the effect on indirect heat accounting methods due to the different number of radiator elements [12].

### *3.2. Test conditions*

The hydraulic distribution considered for the tests is the main risers vertical distribution, as it represents the most common case for the application of indirect heat accounting systems and one of the most used hydraulic distribution layouts of centralized heating systems in the European building stock.

A test plan of a total duration of 23 days has been designed in order to reproduce the typical operating conditions of indirect heat accounting systems; in particular, concerning the heater control, the following working conditions have been considered (each one applied for the 50% of the total test duration):

- operation at constant forward temperature from the heater of about 55 °C;
- operation at variable forward temperature from the heater to mimic the response of a heating curve based on the outdoor temperature (climatic temperature control): namely, according to a linear curve, a minimum heater supply temperature of 40 °C is set for an outdoor temperature of 20 °C, and a maximum heater forward temperature of 70 °C is set with 0 °C outdoor temperature.

Furthermore, to evaluate the influence on heat accounting due to radiator flow regulation, the following control strategies of STVs have been considered for both test periods at constant and variable heater forward temperature:

- Room temperature Set-Point (RSP): the STV opening is controlled as a function of the difference between the measured room temperature and a given temperature set-point. The RSP values consist of day and night temperature set-points, which are increased at higher floors to take into account the indoor temperature stratification during tests.
- Valve Position Set-Point (PSP): the STV opening follows a programmed daily pattern (i.e. constant percentage opening over several hours).

The automatic ventilation system has been used to control the indoor temperature and to limit the temperature stratification inside the laboratory; indoor air temperature is kept within (22 ± 3) °C during the whole test duration.

It is important to observe that the test period begins and ends with the heater off, so that all the radiator surfaces are in thermal equilibrium at the indoor ambient temperature both at the beginning and at the end of the test. This is required for a proper validation of indirect estimations of radiator heat outputs by means of reference direct thermal energy measurements. To this purpose, an adhesive PT100 thermometer, fixed on the surface of one of the cast iron radiators (as they are characterized by the longest heating/cooling transients) next to the HCA mounting position, has been used to control the complete cooling down of radiator surfaces and ascertain the end of the test.

Because of the limited duration of the test (23 days), all HCAs have been arbitrarily set with the maximum allowable value for the configuration parameter (equal to 999), in order to force them to the highest counting rate and increase their resolution. Details on the uncertainty contribution associated to HCA resolution are given in Section 1.A.2. In order to calculate the share of heat consumption, HCAs readings are re-scaled by the actual configuration parameters defined by the EN 834 Standard and determined from radiator manufacturers' data shown in Table 1. In the case of the application of the improved heat accounting method to HCAs, the elements of matrix $\boldsymbol{A}$ are obtained dividing the HCA readings by the maximum arbitrary value of the configuration parameter, and the share of heat consumption is evaluated by re-scaling the HCA readings by means of the estimated parameters $\hat{\theta}_j$.

Finally, concerning the application of the method both to HCAs and to STVs, it must be noted that the measurement of the overall thermal energy exchanged by the heat transfer fluid is calculated as the sum of the thermal energies measured by the reference radiator DHMs, which is the same as considering an ideal thermal-hydraulic distribution with null heat losses through the pipework. In order to evaluate the influence of heat losses on the accuracy of the method, the sum of radiator thermal energy measurements has been increased according to different fractions of heat loss through the pipework, analysing the response of the method at different levels of thermal insulation of the heating system distribution.

### *3.3. Definition of measurands and performance indicators*

The reference measurement of the thermal energy $Q_{ref}$ exchanged by the water flowing through a radiator is provided by radiator DHMs, and is obtained by the following model:

$$Q_{ref} = \int \left(\rho \dot{V} c_p \Delta T + q_{co}\right) dt \tag{11}$$

where $\rho$ and $c_p$ are, respectively, the density and the specific heat capacity of water, $\Delta T$ is the temperature difference between the inlet and outlet flow section of the radiator, $\dot{V}$ the radiator volumetric flow rate and $t$ is the time. The term $q_{co}$ is the thermal power correction that takes into account possible radiator flow rates lower than the low flow cut-off limit of radiator flow meters, set to 2.5 L h$^{-1}$. Such a correction is needed since the operation of STVs, in particular when they are programmed with room temperature set-points, may determine very low radiator flow rates, which cannot be detected by radiator flow meters. The thermal power correction is calculated as follows:

$$q_{co} = \begin{cases} \frac{1}{2}\dot{V}_{co}\,\rho\,c_p\,\Delta T & \text{if } \dot{V} = 0 \text{ for more than 1 hour } AND\ \ \Delta\text{T} \geq 8\text{ K} \\ 0 & \text{otherwise} \end{cases} \tag{12}$$

where $\dot{V}_{co}$ is the cut-off flow rate of radiator flow meters.

As described by equation (12), the correction is non-zero only if two conditions are simultaneously satisfied: the flow rate given by the radiator flow meter is zero for more than 1 hour and the radiator inlet-outlet temperature difference measurement is higher than 8 K. Based on the types of radiator installed at the INRIM central heating system mock-up, the first condition takes into account that the mean time needed by radiator surfaces to cool down to indoor ambient temperature is approximately 1 hour, whereas the second condition set the minimum inlet-outlet temperature difference that can be associated to a radiator flow rate lower than 2.5 L $h^{-1}$. The correction has been considered as uniformly distributed between zero and $2{\cdot}q_{co}$ as described in Section 1.A.1.

The accuracy of indirect heat accounting systems is evaluated comparing the fractions of heat consumption of each radiator or subset of radiators (share of heat consumption) against the reference fractions obtained by means of radiators DHMs. The fractions of heat consumption, expressed in %, are defined as follows:

$$f_{X,s} = 100\,\frac{X_s}{\sum_{s=1}^{N_s} X_s} \tag{13}$$

where $X_s$ represents the heat consumption related to the *s*-th subset of radiators, which can be determined either by the direct thermal energy measurements provided by reference radiator DHMs, or by the corresponding estimations of heat consumption or allocation units provided by the indirect heat accounting systems. In equation (13), $N_s$ is the number of radiator subsets, in which the heating system is grouped, and can range from 2 to $K$, where $K$ is the overall number of radiators considered for the assessment of the share of heat consumption. It is worth observing that the assessment of the quantities $X_s$ is based on the preliminary definition of radiator subsets, to which the individual heat consumptions are related. Namely,

$$X_s = \sum_{j=1}^{N} X_{j,s} \tag{14}$$

where $X_{j,s}$ is the heat consumption of the *j*-th radiator of the *s*-th subset and $N$ is the number of radiators of the *s*-th subset, which can range from 1 to ($K$ – 1).

The errors associated to the estimations of the fractions of heat consumption, or heat allocation errors, are calculated as:

$$E_{f,X,s} = f_{X,s} - f_{ref,s} \tag{15}$$

where $f_{ref,s}$ is the reference fraction of heat consumption of the *s*-th subset of radiators, which is calculated from the direct thermal energy measurements provided by reference radiator DHMs. According to the definition of the fraction of heat consumption, the heat allocation errors $E_{f,X,s}$ are normalized with respect to the overall heat consumption of all the radiator subsets, and their corresponding mean value is equal to zero. In other words, they represent the errors of the heat accounting system in terms of individual fractions of the total heat consumption and can be used to compare different indirect heat accounting systems.

The performance of indirect heat accounting systems can be also described by the following global indicators:

- standard deviation of heat allocation errors $\sigma\left(E_{f,X,s}\right)$;
- maximum and minimum of $E_{f,X,s}$;
- Mean Absolute Percentage Error, $MAPE_X$, defined as:

$$MAPE_X = \frac{100}{N_s}\sum_{s=1}^{N_s} \frac{\left|E_{f,X,s}\right|}{f_{ref,s}} \tag{16}$$

The reason to include the *MAPE* is to provide the estimate of the average percentage error associated to the bill of individual thermal energy consumption, since the *MAPE* normalizes the heat allocation errors by the corresponding fractions of heat consumption of radiator subsets.

All these global indicators give a measure of the fairness of the heat allocation, describing the distribution of the errors associated to the estimated fractions of heat consumption. It is important to observe that in the case of the share of heat consumption among groups or subsets consisting of more than one radiator, the errors associated to the estimations of single radiator heat consumptions cumulate into the estimated thermal energy consumption of the corresponding group.

In order to compare the performance of the improved heat accounting method (applied both to HCAs and to STVs), against that of the HCA system (at nominal parameters configuration), it is also useful to define the following global indicators, which give a measure of the effectiveness of the method:

- Global deviation of heat allocation errors with respect to HCAs, $\Delta E_{HCA,X}$, defined as:

$$\Delta E_{HCA,X} = \sum_{s=1}^{N_s}\left(\left|E_{f,X,s}\right| - \left|E_{f,HCA,s}\right|\right) \tag{17}$$

- Percentage $P_L$ of radiator subsets, whose absolute values of heat allocation errors resulting from the application of the improved heat accounting method are lower than the corresponding absolute values of errors obtained by HCAs.

The global uncertainty $u_{G,X}$ associated to the average individual fraction of heat consumption can be calculated as follows:

$$u_{G,X} = \sqrt{\sigma^2\left(E_{f,X,s}\right) + \bar{u}^2\left(E_{f,X,s}\right)} \tag{18}$$

where $\bar{u}(E_{f,X,s})$ is the average standard uncertainty associated to the heat allocation errors, which is calculated considering the heat allocation errors of radiator subsets as fully correlated. The details of the uncertainty analysis are given in Appendix A.

## 4. Results and discussion

In this section, the results of the experimental activity carried out at the INRIM centralized heating system mock-up, and the outcomes of preliminary pilot test performed in a building in Neuchatel (2018-2019 heating season), Switzerland, are presented and discussed. Concerning the pilot test at the Neuchatel building, the results are related to the application of the method to STVs; the main objective was to evaluate the performances of the SMINTEBI heat accounting system in a real environment, under typical user and weather influences.

### *4.1. Results of tests at the INRIM centralized heating system mock-up*

At first, the improved heat accounting method has been tested on a typical HCA system, which represents the most used indirect heat accounting system on the market. For this case, the influence of the uncertainty sources associated to transient heat transfer conditions and heat losses through the pipework, along with the sensitivity to input uncertainties on the prior knowledge $\boldsymbol{\theta_0}$ of radiator parameters, have been analysed in Section 4.1.1.1.

Then, the improved heat accounting method has been applied to STVs, and the heat allocation results for both the applications to HCAs and STVs have been validated by comparison against the share of heat consumption obtained by the reference radiator DHMs. In addition, the shares of heat consumption provided by the improved heat accounting method, for both the HCAs and the STVs cases, have been compared against those obtained by the typical HCA system. In particular, the results of the experimental validation have been expressed in terms of the global indicators of performance defined in Section 3.3, which allow comparing the accuracy of the new method against that of the typical HCA system at nominal parameters configuration. The results of the experimental campaign are shown considering two different scenarios. In the first one, described in Section 4.1.1, the fractions of heat consumption are evaluated for each radiator, whereas in the second one, described in Section 4.1.2, the individual thermal energy consumptions are related to 8 subsets of radiator or "virtual apartments". Fig. 5 shows a picture of the indirect heat accounting systems under test at the INRIM centralized heating system mock-up.

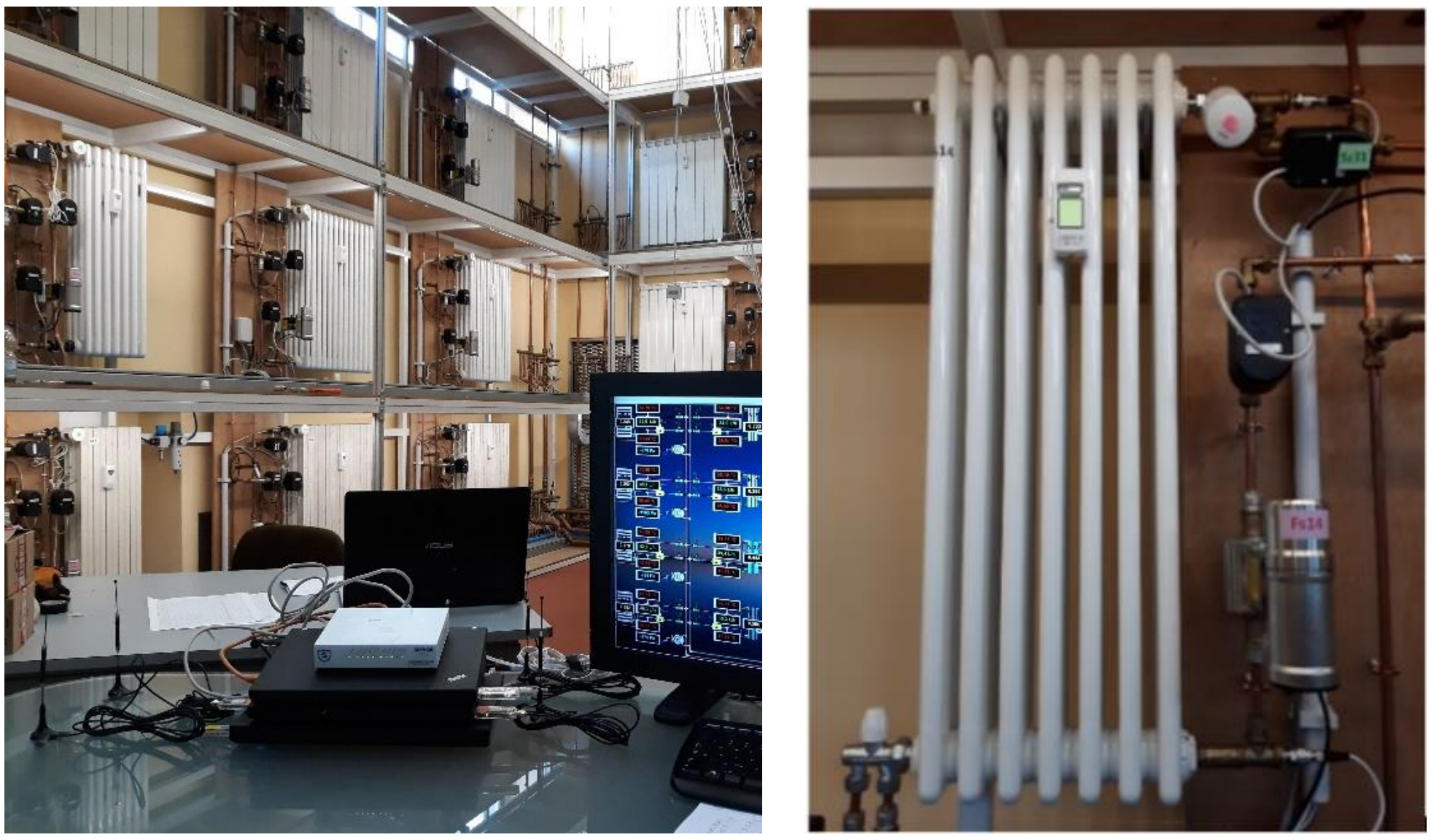

**(a)** **(b)**

**Fig. 5.** (a) Indirect heat accounting systems under test at the INRIM centralized heating system mock-up; (b) detail of a radiator equipped with reference DHM (flow meter and inlet-outlet temperature sensors pair), HCA and STV.

It must be observed that two radiators of the system, i.e. radiators S18 (10 elements cast iron radiator) and S19 (5 elements cast iron radiator), had to be discarded from the analyses, because of a malfunction of the corresponding STV inlet temperature sensors, which showed an error of up to 7 °C. This was discovered just after the experimental campaign, when all the STV valves were put in a climatic chamber at CSEM to check their temperature measurement accuracy.

### *4.1.1. Radiator based heat allocation*

The results of this scenario are obtained considering that the individual heat consumptions correspond to single radiator heat consumptions, i.e. each radiator of the system is regarded as well as a single flat of an apartment building. The analysis of the share of heat consumption over the maximum number of radiator subsets, in which the heating system can be grouped, allows evaluating the performance of indirect heat accounting systems, excluding possible combinations of errors associated to different estimations of radiator heat consumption.

Before comparing the accuracies of indirect heat accounting systems, it is important to analyse the sensitivity of the improved heat accounting method to transient heat transfer conditions (due to the thermal inertia of the system), heat losses through the pipework and input uncertainties on the prior knowledge of radiator parameters. This sensitivity analysis concerns the improved heat accounting method applied to HCAs (hereinafter referred to as “improved HCA method”), in particular:

- The influence of transient heat transfer conditions is analysed by comparing the results obtained considering different frequencies in the acquisition of HCAs and DHM readings, which means different durations of the integration period for HCAs and DHM readings, and different numbers of energy samplings used for parameters estimation. Constant sampling frequencies ranging from 0.33 $h^{-1}$ down to 0.03 $h^{-1}$ (i.e. a number of energy samplings ranging from about 180 to 18 respectively) have been analysed. The analysis is carried out considering a null heat loss contribution through the pipework (overall heat consumption obtained as the sum of radiator DHMs measurements).
- The effect of the heat losses through the pipework is analysed considering the overall thermal energy exchanged by the heat transfer fluid, as given either by the sum of the measurements provided by the radiator DHMs, equal to 7746.15 kWh (null contribution due to the heat loss through the pipework), or by the sum of radiator DHMs measurements increased by different heat loss percentages (heat losses from 5% to 20%). The analysis is carried out considering the maximum frequency in the acquisition of HCAs and DHMs readings. It is worth observing that the actual heat loss through the pipework of the centralized heating system mock-up during tests, was around 20% of the total heat consumption measured by a DHM installed at the main flow section of the hot water distribution.
- The effect of input uncertainties on the prior knowledge $\boldsymbol{\theta_0}$ of radiator parameters is analysed considering both constant percentage offsets, and uniformly distributed relative errors on initial guess parameters $\boldsymbol{\theta_0}$. The input uncertainties are estimated from the possible configuration errors that may happen during the setup of indirect heat accounting devices (e.g. configuration of heat accounting devices installed on radiators whose nominal thermal model parameters are not available). The analysis is carried out considering a null heat loss contribution through the pipework and the maximum frequency in the acquisition of HCAs and DHMs readings.

#### *4.1.1.1. Sensitivity analysis*

Concerning the analysis of the influence of transient heat transfer conditions, Table 2 shows the values of the regularization parameter $\lambda$ determined by the L-curve method, and associated to the measurement data sets gathered with different sampling frequencies $\varphi$ for HCAs and DHMs acquisition. These parameters give a measure of how ill-posed the problem is; it is important to observe that the higher the sampling frequency, the higher the expected influence of thermal inertia, but, on the other hand, the higher the size of matrix $\boldsymbol{A}$. Conversely, with lower sampling frequencies, the influence of transient heat transfer conditions is expected to decrease, but also the size of matrix $\boldsymbol{A}$ decreases, because of the lower number of energy samplings available for parameters estimation.

**Table 2.** Values of regularization parameters $\lambda$ associated to the batches of data gathered with different sampling frequencies $\varphi$ for HCAs and DHMs acquisition.

| $\varphi$ / h$^{-1}$ | $\lambda$ |
|---|---|
| 0.33 | 0.0240 |
| 0.16 | 0.0230 |
| 0.10 | 0.0440 |
| 0.07 | 0.0651 |
| 0.03 | 0.0430 |

Table 3 shows the global indicators of performance of the improved HCA method, obtained considering the different sampling frequencies reported in Table 2. The method responds well to the decrease of the acquisition frequency, and consequently to the reduction of the number of energy samplings, with limited increase of standard deviations of errors $\sigma(E_f)$ and negative values of the global deviation of heat allocation errors with respect to HCAs ($\Delta E_{HCA}$< 0), down to $\varphi$= 0.07 h$^{-1}$. Moreover, the impact of transient heat transfer conditions at high sampling frequencies seems to be mitigated by the corresponding high number of energy samplings. The increase of $MAPE$ values with decreasing sampling frequency is due to the slight increase of heat allocation errors, especially for radiators characterized by lower heat consumptions.

**Table 3.** Influence of transient heat transfer conditions and sampling frequency on the improved HCA method, at null heat loss through the pipework.

| Global indicators | Sampling frequency of HCAs and DHMs acquisition, $\varphi$ / h$^{-1}$ | | | | |
|---|---|---|---|---|---|
| | 0.33 | 0.16 | 0.10 | 0.07 | 0.03 |
| $\sigma(E_f)$ / % | 0.17 | 0.19 | 0.18 | 0.19 | 0.21 |
| $max(E_f)$ / % | 0.25 | 0.31 | 0.26 | 0.25 | 0.33 |
| $min(E_f)$ / % | -0.45 | -0.46 | -0.44 | -0.45 | -0.50 |
| $MAPE$ / % | 5.74 | 6.46 | 6.22 | 6.35 | 7.14 |
| $P_L$ / % | 53 | 50 | 50 | 47 | 47 |
| $\Delta E_{HCA}$ / % | -0.91 | -0.36 | -0.50 | -0.34 | 0.20 |

Fig. 6 shows the comparison between the heat allocation errors $E_f$ obtained for each radiator by the improved HCA method, and the heat allocation errors $E_{f,HCA}$ obtained by HCAs at nominal parameters configuration. The results are plotted for the different sampling frequencies considered in Table 3. The comparison is expressed in terms of the absolute value of the ratio $E_f/E_{f,HCA}$ as a function of $E_{f,HCA}$ and demonstrates the capability of the method to reduce the higher heat allocation errors obtained by HCAs at nominal parameters configuration, especially those characterized by absolute values higher than 0.25%.

On the other hand, a percentage of radiators equal to $100 - P_L$ shows increased heat allocation errors compared to HCAs; however, since these radiators are characterized by the lowest values of $E_{f,HCA}$, the increments (which are inversely proportional to $E_{f,HCA}$) are such that the global deviation of heat allocation errors with respect to HCAs is negative ($\Delta E_{HCA} < 0$), and the absolute values of the maximum and minimum heat allocation errors are lower than the ones of HCAs (except for $\varphi = 0.03\ h^{-1}$).

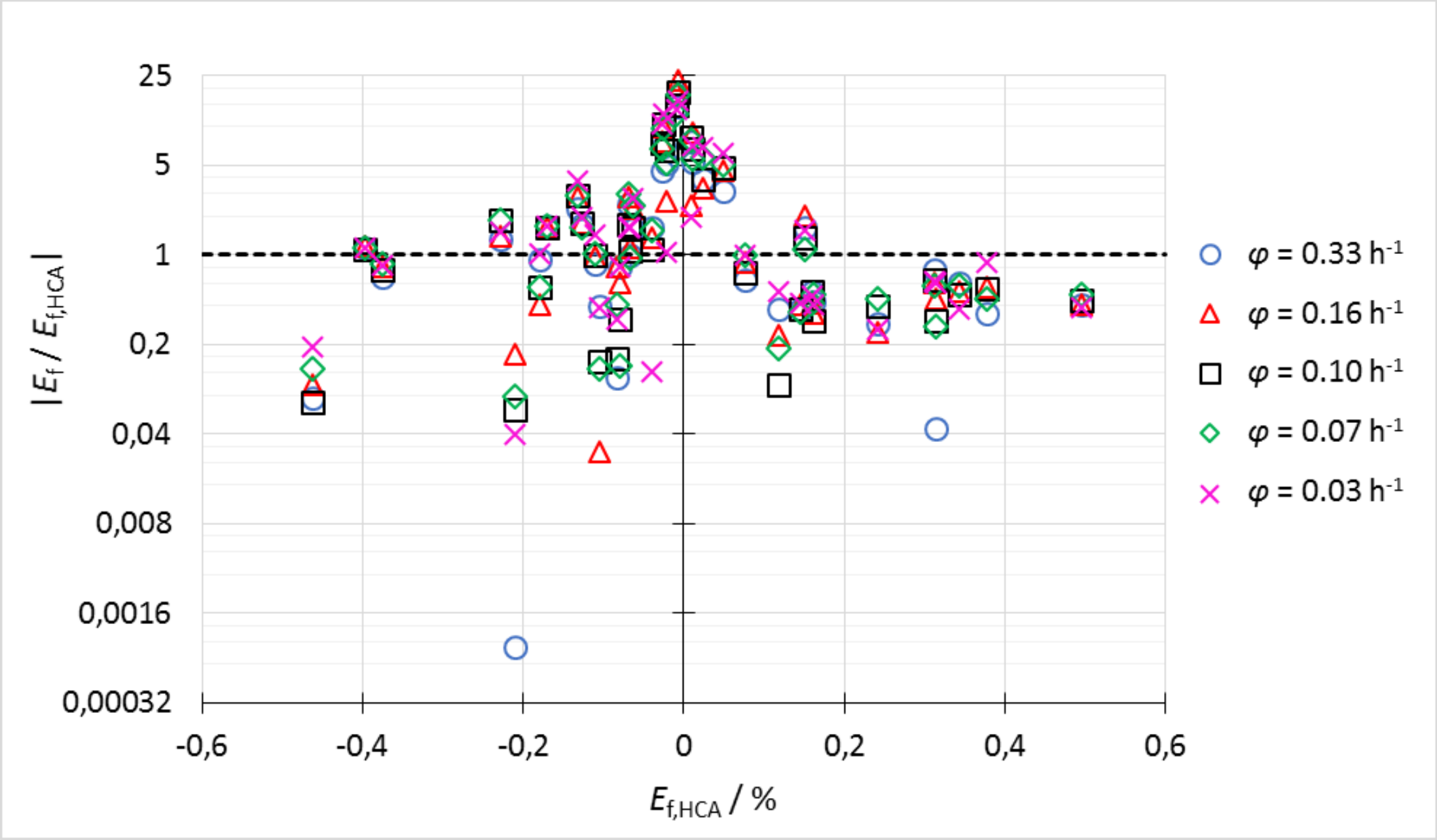

**Fig. 6.** Heat allocation errors obtained by the improved HCA method, compared to those of HCAs at nominal parameters configuration; results are plotted for the different sampling frequencies $\varphi$.

Concerning the influence of the heat losses through the pipework, Table 4 shows the results obtained for the improved HCA method, considering different fractions of heat loss, from 0% (ideal thermal hydraulic distribution with null heat loss through the pipework) to 20%. In particular, for each case, the overall thermal energy measurements associated to the different energy samplings are obtained increasing the sum of radiator DHMs measurements by the fraction of heat loss through the pipework. The results in Table 4 show a worsening of all global indicators with increasing heat losses, indicating that the method is expected to work properly in centralized heating systems characterized by a good thermal insulation of the pipework (heat losses preferably lower than 10%). Such a sensitivity of the method with respect to heat losses can be mitigated including in equation (6) a proper estimation of the thermal energy dissipated through the pipework. In particular, the overall thermal energy used as input for the model, should be considered as equal to the fraction of voluntary heat consumption, which is usually determined to split the energy bill in a variable or voluntary part (which is shared among the tenants according to their individual heat consumptions and does not account for heat losses through the pipework), and a fixed part (based on building envelope and thermal insulation characteristics and shared among the tenants according to the size of their apartments). The fraction of voluntary heat consumption should be evaluated with an uncertainty preferably lower than 10%.

**Table 4.** Influence of heat losses through the pipework on the improved HCA method.

| Global indicators | Heat losses through the pipework | | | |
|---|---|---|---|---|
| | 0% | 5% | 10% | 20% |
| $\sigma(E_f)$ / % | 0.17 | 0.19 | 0.22 | 0.28 |
| $max(E_f)$ / % | 0.25 | 0.27 | 0.36 | 0.54 |
| $min(E_f)$ / % | -0.45 | -0.49 | -0.51 | -0.57 |
| $MAPE$ / % | 5.74 | 6.83 | 7.88 | 10.44 |
| $P_L$ / % | 53 | 42 | 37 | 29 |
| $\Delta E_{HCA}$ / % | -0.91 | -0.04 | 0.84 | 2.92 |

Fig. 7 shows the relative deviations between the estimated radiator parameters $\hat{\boldsymbol{\theta}}$ and their prior knowledge or initial guess $\boldsymbol{\theta}_0$, obtained for the ideal case of null heat losses through the pipework. It is worth observing that in the improved HCA method, the prior knowledge $\boldsymbol{\theta}_0$ consists of the nominal parameters used for the typical configuration of HCAs, which are based on EN 834 Standard and radiators manufacturers' data. The deviations are plotted for each type of radiator, as functions of the corresponding EN 442 nominal thermal power.

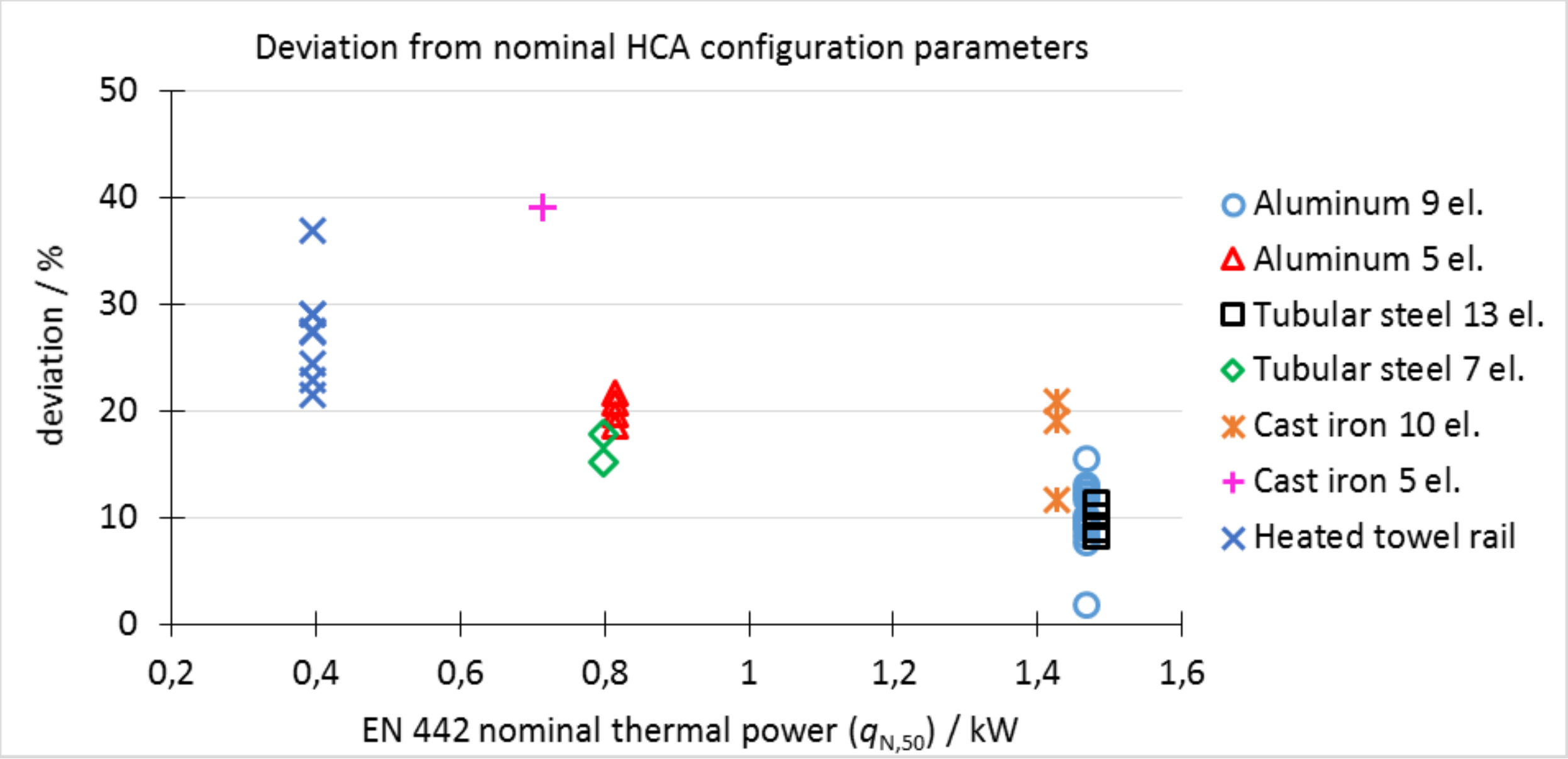

**Fig. 7.** Relative deviations between estimated radiator parameters and their prior knowledge; the deviations are plotted for each type of radiator, as functions of the corresponding EN 442 nominal thermal power.

The deviations of the estimated radiator parameters with respect to the corresponding prior knowledge mainly depend on the particular installation and operating conditions in which radiators are tested. Moreover, the differences between the estimates associated to radiators of the same type, are due to the different working conditions occurring at the test facility, where radiators may be characterized by different heat transfer conditions depending on their position in the thermal-hydraulic circuit. For instance, as detailed hereinafter in Section 4.1.2 concerning the virtual apartments scenario, radiators placed at the first floor of the hydraulic distribution layout, especially those on the north wall, which are closer to the air intakes of the ventilation system, are characterized by higher deviations of their estimated parameters with respect to the prior knowledge. This can be explained by the underestimation associated to the initial guess of radiator parameters (valid for nominal heat transfer conditions according to EN 442 and EN 834 Standards), with respect to the enhanced convective heat transfer conditions, which these radiators are actually exposed to. From Fig. 7, it is also interesting to observe that for tubular steel, aluminium, and cast iron sectional radiators, the lower the number of elements, the higher the deviation. This also agrees with the fact that the overall heat transfer at the radiator surface is expected to decrease as the number of elements increases [12].

The sensitivity of the improved HCA method to input uncertainties on the prior knowledge $\boldsymbol{\theta_0}$ of radiator parameters has been analysed considering constant relative offsets and uniformly distributed relative errors applied to the elements of vector $\boldsymbol{\theta_0}$. The results of the analysis are reported in Table 5 in terms of global indicators of performance. It can be observed that the method has a good response to uniformly distributed relative errors on the prior knowledge, providing negative values of the global deviation of heat allocation errors $\Delta E_{HCA}$ (the global indicators are calculated as mean values of the results obtained considering five different vectors $\boldsymbol{\theta_0}$, whose elements are uniformly distributed within ±10% and ±20% of the nominal configuration parameters of HCAs). On the other hand, the method has a worse response to constant relative offsets on $\boldsymbol{\theta_0}$, particularly when the offset shifts the initial guess of radiator parameters too far from the actual solution; this is the case of -20%, -10%, and +20% offsets, which entail positive values of $\Delta E_{HCA}$, i.e. increased heat allocation errors with respect to HCAs at nominal parameters configuration. However, constant relative offsets on $\boldsymbol{\theta_0}$ may also have positive effects on heat allocation if they move the initial guess close to the actual solution; this is the case of an offset of +10%, which approximately corresponds to the mean relative deviation between $\widehat{\boldsymbol{\theta}}$ and $\boldsymbol{\theta_0}$ observed in Fig. 7 for the radiators characterized by the higher nominal thermal power (about 50% of the total number of radiators).

**Table 5.** Sensitivity of the improved HCA method to input uncertainties on the prior knowledge of radiator parameters.

| Global indicators | constant relative offset on $\boldsymbol{\theta_0}$ | | | | uniformly distributed relative errors on $\boldsymbol{\theta_0}$ | |
|---|---|---|---|---|---|---|
| | +10% | -10% | +20% | -20% | ±10% | ±20% |
| $\sigma(E_f)$ / % | 0.17 | 0.23 | 0.21 | 0.31 | 0.22 | 0.33 |
| $max(E_f)$ / % | 0.35 | 0.38 | 0.50 | 0.64 | 0.49 | 0.74 |
| $min(E_f)$ / % | -0.39 | -0.52 | -0.47 | -0.60 | -0.50 | -0.77 |
| $MAPE$ / % | 5.14 | 8.17 | 6.44 | 11.66 | 6.70 | 9.84 |
| $P_L$ / % | 74 | 37 | 42 | 29 | 58 | 55 |
| $\Delta E_{HCA}$ / % | -1.15 | 1.07 | 0.04 | 3.95 | -1.53 | -1.62 |

#### *4.1.1.2. Comparison between indirect heat accounting systems*

Concerning the comparison between the accuracies of the indirect heat accounting systems under test, the results of the improved heat accounting method applied both to HCAs and to STVs are obtained considering null heat loss through the pipework, i.e. overall thermal energy measurements calculated as the sum of thermal energy measurements provided by reference radiators DHMs. Table 6 shows the global indicators of performance obtained for the three indirect heat accounting systems under test, namely HCA at nominal parameters configuration (referred to as “HCA”), the improved HCA method, and the improved heat accounting method applied to STVs (referred to as “SMINTEBI”).

The results show an improvement of all the global indicators of the improved HCA method with respect to the HCA system, in particular the reduction of the standard deviation of errors and a 10% reduction of the *MAPE* indicate a fairer heat allocation with respect to the HCA system. Moreover, a global deviation of heat allocation errors $\Delta E_{HCA}$ of -0.91%, characterized by a global uncertainty $u(\Delta E_{HCA})$ of about 0.30% (calculated as the square root of the sum of the squared global uncertainties $u_G^2$ associated to HCA and improved HCA method), demonstrates the effectiveness of the method in reducing the errors associated to the HCA system. The reduction of errors observed for the improved HCA method, is due to its capability to calibrate radiator parameters at the actual test site conditions, outperforming typical HCA systems, which rely on standard parameters obtained in controlled conditions (reference installation and heat transfer conditions defined by the EN 442 Standard).

The SMINTEBI system shows higher score in the global indicators with respect to HCAs, caused almost exclusively by the high heat allocation error associated to one of the radiators with the lowest heat consumption (radiator S7, characterized by a fraction of heat consumption $f_{ref}$ of 0.64%). Such a high heat allocation error, equal to 1.23%, is likely due to the use of the inlet fluid temperature measurement provided by the STV, as an input for the model described by equation (10) . In fact, if a radiator is characterized by most of operation at low-partial[1] STV opening (like radiator S7), the model tends to overestimate the temperature difference between the heat transfer fluid and the surrounding environment, being unable to clearly discriminate between partial and full STV openings.

**Table 6.** Comparison between the global indicators of performance of the three indirect heat accounting systems under test for the radiators heat allocation scenario.

| Global indicators | Radiators heat allocation scenario | | |
|---|---|---|---|
| | HCA | Improved HCA | SMINTEBI |
| $\sigma(E_f)$ / % | 0.21 | 0.17 | 0.33 |
| $max(E_f)$ / % | 0.49 | 0.25 | 1.23 |
| $min(E_f)$ / % | -0.46 | -0.45 | -0.58 |
| $MAPE$ / % | 6.40 | 5.74 | 13.49 |
| $P_L$ / % | - | 53 | 42 |
| $\Delta E_{HCA}$ / % | - | -0.91 | 2.52 |
| $\bar{u}(E_f)$ / % | 0.08 | 0.05 | 0.05 |
| $u_G$ / % | 0.22 | 0.18 | 0.33 |

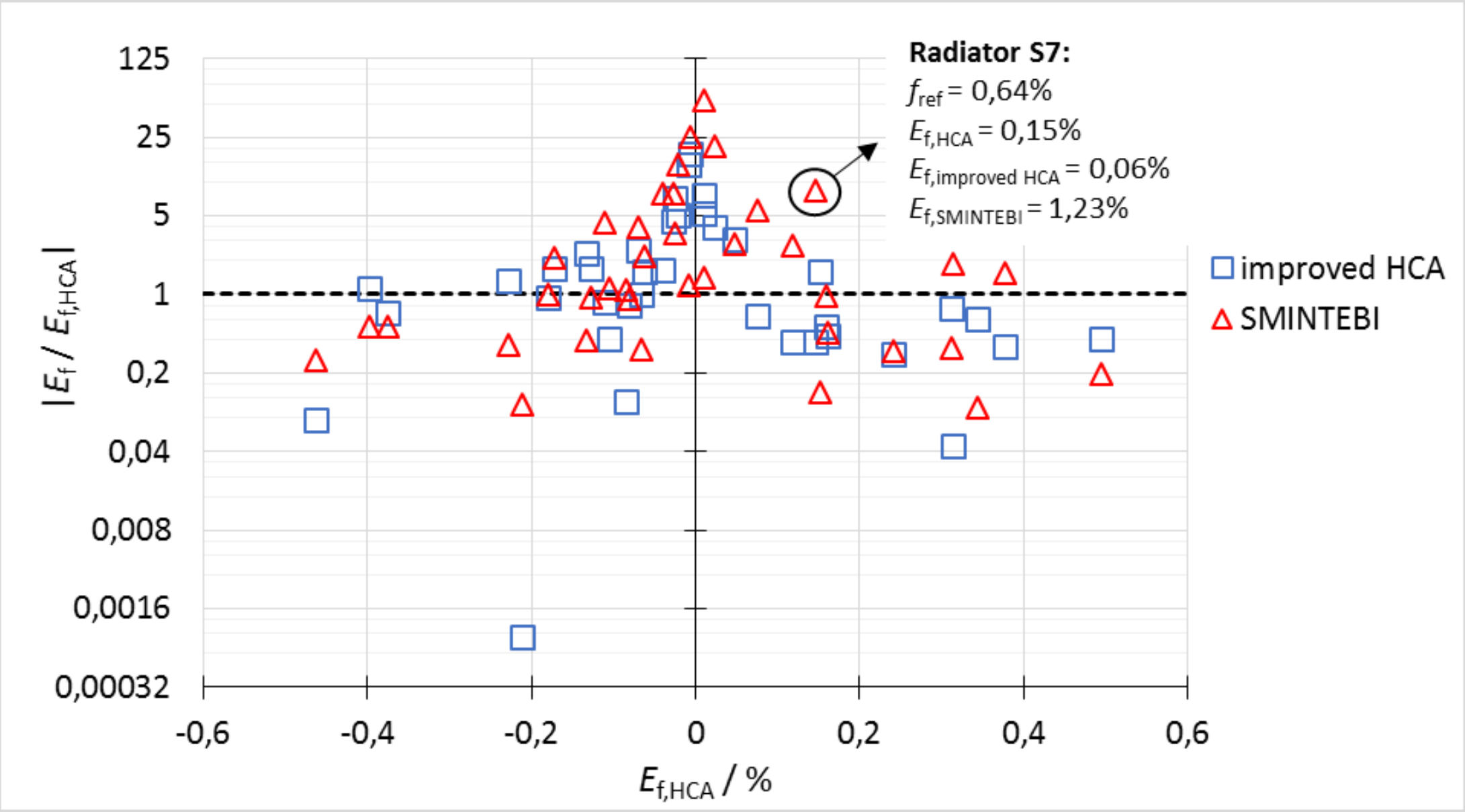

**Fig. 8.** Comparison of heat allocation errors obtained by the improved HCA method and the SMINTEBI system, against those of HCAs at nominal parameters configuration.

[1] Valve opening close to 0%, resulting in a low flow and thus a low fraction of heat consumption.

Fig. 8 shows the comparison between the heat allocation errors obtained by the improved HCA method and the SMINTEBI system, against those obtained by the typical HCA system, expressed in terms of the absolute values of the ratios $E_f/E_{f,HCA}$ as functions of $E_{f,HCA}$. In particular, despite the point related to the results for radiator S7, the capability of the SMINTEBI system to reduce the higher heat allocation errors of HCAs can be observed.

### *4.1.2. Apartments heat allocation scenario*

In order to evaluate the performance of the indirect heat accounting systems in a more realistic scenario, the radiators have been grouped per wall and floor to have 8 virtual apartments of 5 radiators each, as shown in Fig. 9. Each apartment consists of different types of radiators with different thermal outputs, as well as in most of real applications.

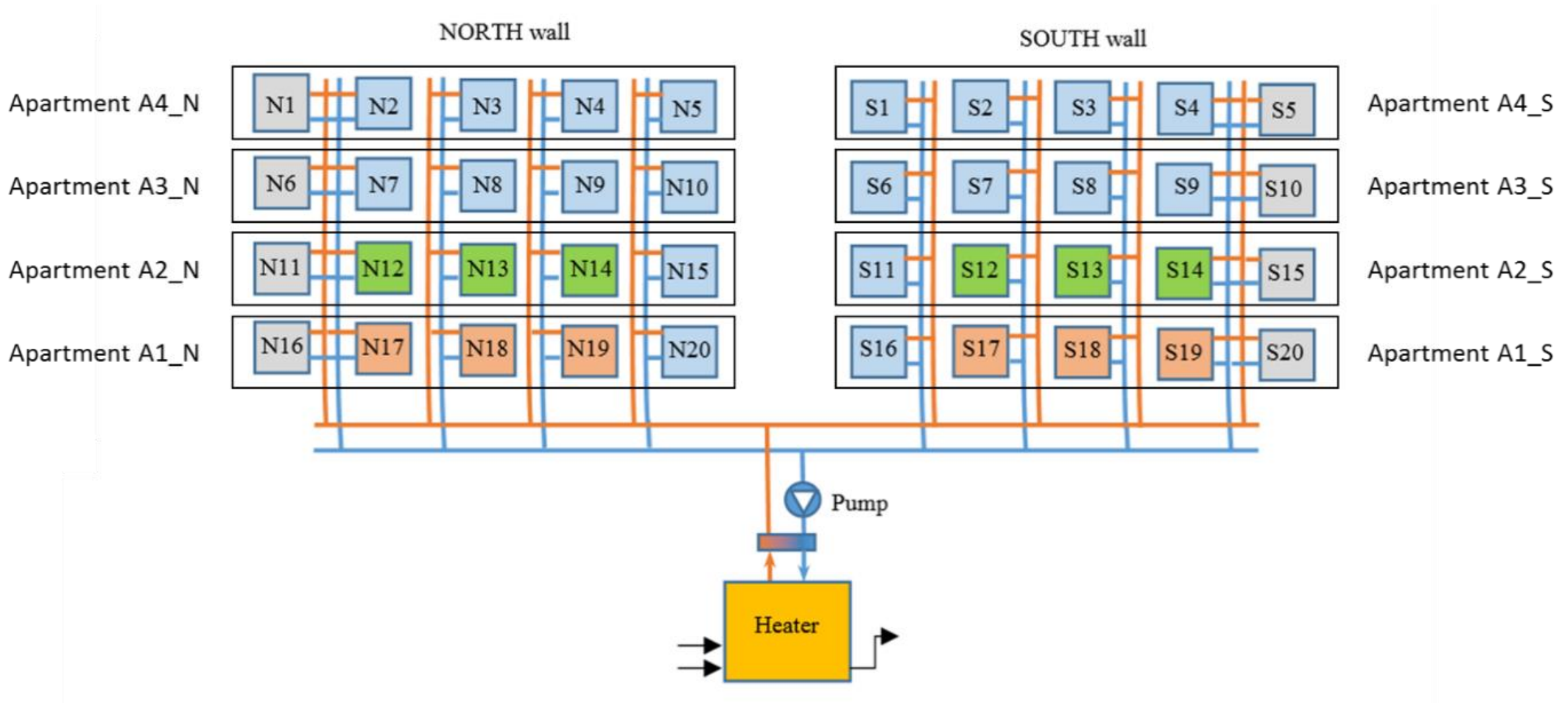

**Fig. 9.** Identification of the 8 virtual apartments at the INRIM centralized heating system mock-up.

In this situation, possible combinations of errors associated to the estimations of radiators heat consumptions can occur, leading to particularly high deviations if the virtual apartments consist of radiators characterized by similar off-design installation and heat transfer conditions. This is the case of the first floor virtual apartments, especially A1_N, whose radiators are exposed to enhanced convective heat transfer conditions due to the proximity of the air intakes of the ventilation system (particularly concerning radiator N17). Concerning these radiators, the typical HCA configuration based on the EN 834 Standard and the radiator manufacturer's data (nominal configuration parameters) is expected to provide underestimated heat consumption results. On the other hand, the improved heat accounting method allows determining the characteristic thermal model parameters of radiators according to the actual heat transfer conditions and installation effects, to which these radiators are exposed.

Concerning the improved HCA method, Table 7 shows, for each apartment, the ratio $\gamma$ between the sum of estimated radiators parameters and the corresponding sum of the prior knowledge of radiator parameters (based on the nominal configuration parameters available from manufacturers' data). In addition, the percentage deviations of the ratios $\gamma$ of each apartment with respect to the ratio $\gamma_{TOT}$ calculated over the whole set of radiators, are given. From the results shown in Table 7, it can be observed that the first floor virtual apartments A1_N and A1_S are those characterized by the highest positive percentage deviations of $\gamma$ with respect to $\gamma_{TOT}$. This agrees with the actual heat transfer and installation conditions to which radiators are exposed, particularly as far as apartment A1_N is concerned.

**Table 7.** Comparison between the sum of characteristic radiators parameters estimated by the improved HCA method, and the sum of the corresponding prior knowledge of parameters (equal to the nominal configuration parameters of HCAs), for each virtual apartment.

| Apartment | $\gamma$ | $100 \times \frac{\gamma - \gamma_{TOT}}{\gamma_{TOT}}$ / % |
|---|---|---|
| A1_N | 1.211 | 6.6 |
| A2_N | 1.120 | -1.5 |
| A3_N | 1.142 | 0.5 |
| A4_N | 1.118 | -1.6 |
| A1_S | 1.176 | 3.4 |
| A2_S | 1.124 | -1.1 |
| A3_S | 1.112 | -2.2 |
| A4_S | 1.120 | -1.4 |
| $\gamma_{TOT}$ | 1.137 | |

Fig. 10 shows the heat allocation errors of virtual apartments obtained by the three indirect heat accounting systems under test (HCA, improved HCA, and SMINTEBI), plotted with their corresponding expanded uncertainty bars. Heat allocation errors have been calculated with respect to the reference share of heat consumption among the virtual apartments, obtained by means of the reference radiators DHMs. A significant negative deviation of the heat allocation error of apartment A1_N with respect to the other apartments can be observed for the results obtained by HCAs, which demonstrates an underestimation of the corresponding heat consumption. On the other hand, the improved heat accounting method applied both to HCAs and to STVs has been able to mitigate the negative impact on indirect heat accounting due to off-design installation and heat transfer conditions, providing a more uniform distribution of heat allocation errors among the different virtual apartments. In particular, the heat allocation errors associated with the results provided by the improved HCA method are all within ±0.6%. The slightly higher heat allocation errors of the SMINTEBI system are basically due to the model described by equation (10), which does not allow discriminating between low-partial and full STV openings, as discussed in Section 4.1.1.2.

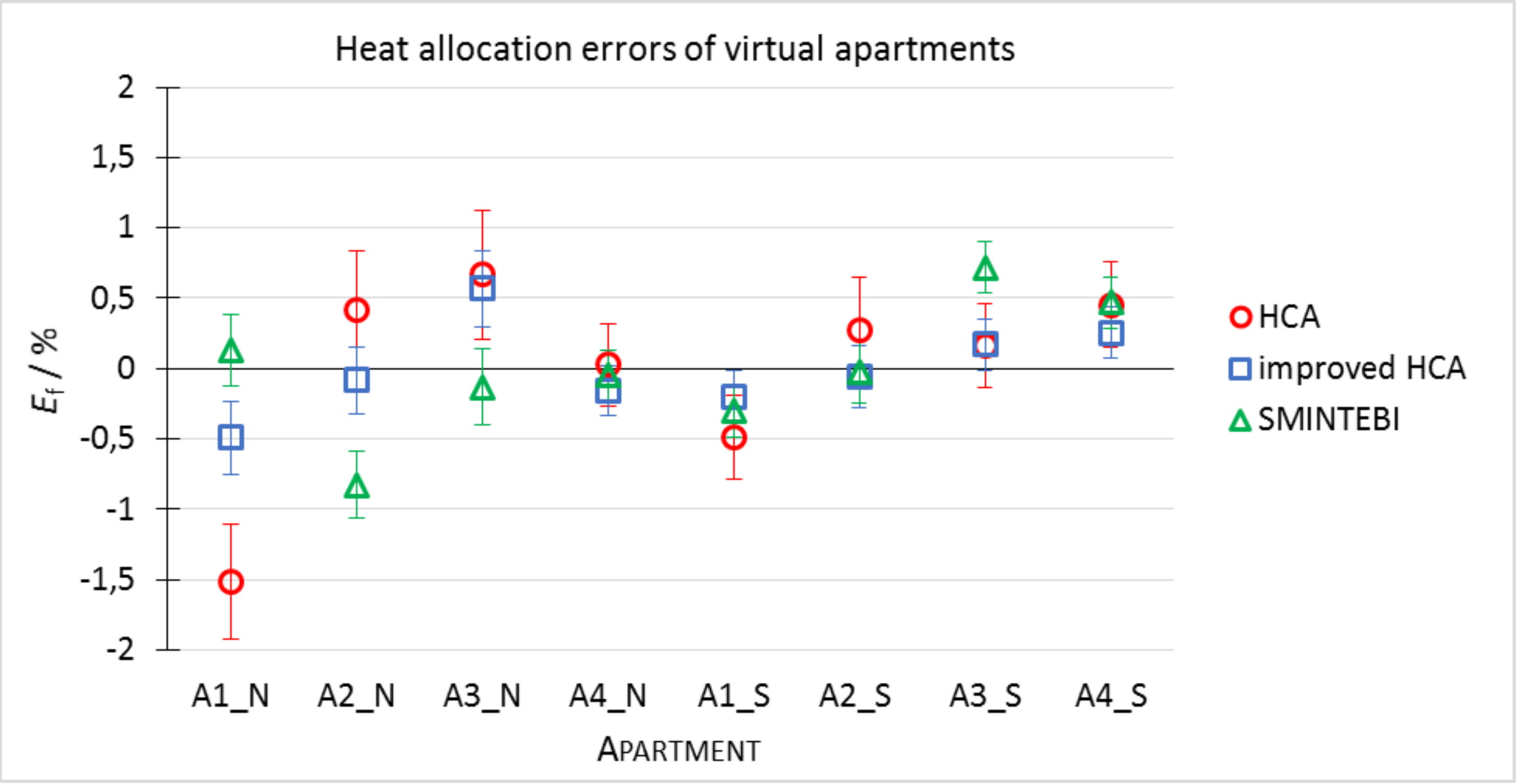

**Fig. 10.** Comparison between the heat allocation errors of virtual apartments, obtained by the three indirect heat accounting systems under test.

Concerning the first floor apartment A1_S, the absolute value of the heat allocation error obtained by the HCA system is lower than the one associated to apartment A1_N despite the similar heat transfer conditions to which radiators are exposed. This is due to two main reasons: first, A1_S consists of only three radiators (S16, S17, and S20), thus the propagation of errors associated to radiators heat consumptions does not entail as a high apartment error as observed for A1_N, and second, A1_S radiators are farther from the air intakes of the ventilation system and less exposed to convective air flows compared to A1_N radiators.

In Table 8, for each virtual apartment, the radiators heat allocation errors are provided, both for the HCA system, and for the improved HCA method. The detail of individual radiators heat allocation errors allows highlighting the effect of the combination of errors on the evaluation of apartments heat allocation, particularly significant for apartment A1_N.

**Table 8.** Comparison between radiators heat allocation errors of each virtual apartment, obtained by the HCA system and the improved HCA method.

| Virtual apartment | Radiators | Radiators heat allocation errors $E_f$ / % | | | | | | | | | |
|---|---|---|---|---|---|---|---|---|---|---|---|
| | | HCA system | | | | | Improved HCA method | | | | |
| A1_N | (N16 → N20) | -0.21 | -0.46 | -0.07 | -0.37 | -0.40 | 0.00 | 0.04 | 0.18 | -0.25 | -0.45 |
| A2_N | (N11 → N15) | -0.04 | -0.07 | 0.34 | 0.31 | -0.13 | 0.06 | -0.07 | 0.21 | 0.01 | -0.30 |
| A3_N | (N6 → N10) | -0.02 | -0.02 | 0.38 | 0.31 | 0.02 | 0.17 | 0.11 | 0.13 | 0.24 | -0.09 |
| A4_N | (N1 → N5) | -0.08 | 0.15 | 0.12 | -0.17 | 0.01 | -0.01 | 0.25 | -0.04 | -0.29 | -0.06 |
| A1_S | (S16, S17, S20) | -0.08 | -0.23 | - | - | -0.18 | -0.07 | -0.30 | - | - | 0.17 |
| A2_S | (S11 → S15) | -0.13 | 0.08 | 0.49 | -0.11 | -0.06 | -0.22 | -0.05 | 0.20 | -0.09 | 0.10 |
| A3_S | (S6 → S10) | -0.01 | 0.15 | 0.16 | -0.10 | -0.03 | -0.12 | 0.06 | 0.07 | 0.04 | 0.12 |
| A4_S | (S1 → S5) | -0.01 | 0.16 | 0.24 | 0.01 | 0.05 | -0.12 | 0.08 | 0.07 | 0.07 | 0.15 |

The comparison between the global indicators of performance of the indirect heat accounting systems under test, obtained for the virtual apartment scenario, is shown in Table 9.

**Table 9.** Comparison between the global indicators of performance of the three indirect heat accounting systems under test for the virtual apartment scenario

| Global indicators | Virtual apartments scenario | | |
|---|---|---|---|
| | HCA | Improved HCA | SMINTEBI |
| $\sigma(E_f)$ / % | 0.70 | 0.32 | 0.47 |
| $max(E_f)$ / % | 0.67 | 0.57 | 0.72 |
| $min(E_f)$ / % | -1.51 | -0.49 | -0.83 |
| $MAPE$ / % | 3.74 | 1.90 | 2.88 |
| $P_L$ / % | - | 75 | 50 |
| $\Delta E_{HCA}$ / % | - | -2.01 | -1.35 |
| $\bar{u}(E_f)$ / % | 0.18 | 0.11 | 0.11 |
| $u_G$ / % | 0.73 | 0.34 | 0.48 |

From Table 9, it can be observed a reduction of the *MAPE* with respect to HCA, i.e. a reduction of the individual billing error, of about 49% for the improved HCA method, and 23% for the SMINTEBI system. Moreover, as observed in Section 4.1.1.2 for the radiators heat allocation scenario, the values of the global deviation of heat allocation errors $\Delta E_{HCA}$ and their associated uncertainties $u(\Delta E_{HCA})$ demonstrate the effectiveness of the method in reducing the errors associated to the HCA system. In particular, the standard uncertainty $u(\Delta E_{HCA})$ is equal to 0.80% for the improved HCA method, and 0.87% for the SMINTEBI system. Finally, the global uncertainty $u_G$ associated to the average individual fraction of heat consumption (that is 12.5% share for each of the 8 apartments) shows a significant reduction compared to the HCA system, i.e. about 50% lower for the improved HCA method, and about 35% lower for the SMINTEBI system.

Concerning the improved HCA method, it is important to observe that heat losses through the pipework have a much lower influence on the accuracy of heat accounting with respect to the radiators heat allocation scenario. In fact, the combination of radiators heat allocation errors, despite their increase due to heat losses through the thermal-hydraulic distribution, allows obtaining an accurate share of heat consumption among the virtual apartments, with negative values of the global deviation of heat allocation errors $\Delta E_{HCA}$ up to 30% heat loss, as shown in Table 10.

**Table 10.** Influence of heat losses on the improved HCA method for the virtual apartments scenario.

| Virtual apartments scenario – Improved HCA method | | | | | |
|---|---|---|---|---|---|
| Global indicators | Heat losses through the pipework | | | | |
| | 0% | 5% | 10% | 20% | 30% |
| $\sigma(E_f)$ / % | 0.32 | 0.29 | 0.30 | 0.40 | 0.53 |
| $max(E_f)$ / % | 0.57 | 0.58 | 0.57 | 0.56 | 0.80 |
| $min(E_f)$ / % | -0.49 | -0.23 | -0.36 | -0.59 | -0.77 |
| $MAPE$ / % | 1.90 | 1.83 | 1.76 | 2.33 | 3.07 |
| $P_L$ / % | 75 | 75 | 75 | 50 | 50 |
| $\Delta E_{HCA}$ / % | -2.01 | -2.11 | -2.20 | -1.49 | -0.70 |
| $\bar{u}(E_f)$ / % | 0.11 | 0.11 | 0.11 | 0.11 | 0.11 |
| $u_G$ / % | 0.34 | 0.31 | 0.32 | 0.41 | 0.54 |

### *4.2. Results of tests of the SMINTEBI system on a real operational site*

In order to assess the operation of the SMINTEBI system in real conditions, a preliminary pilot test was carried out in a building in Neuchatel, Switzerland, over the 2018-2019 heating season. The main objective of this work was to ensure that the proposed method can be deployed in a real environment by showing that it is robust to user and weather disturbances. In addition, it was also important to ensure that the solution could be deployed in a non-ideal environment where the piping and radiators were installed about 20 years ago and the characteristics are not well known.

The site is equipped with a gas condensing boiler, that feeds four individual risers where 41 heat emitters are connected (17 on the 1st riser, 6 on the 2nd riser, 9 on the 3rd riser and 9 on the last riser). The radiators are equipped with STVs and HCAs, in addition heat meters are installed at riser level. The data is acquired continuously by several gateways and sent to a database. The measurements were performed over a period of 21 days.

The results are summarized in Fig. 11 and Table 11. It can be observed that the *MAPE* is reduced from 12.5% (HCA) to 6.7% (SMINTEBI). This improvement is even greater than the one obtained at the INRIM heating system mock-up, showing the capability of the SMINTEBI data driven approach to automatically adapt to test site conditions, outperforming conventional heat allocation methods such as HCA.

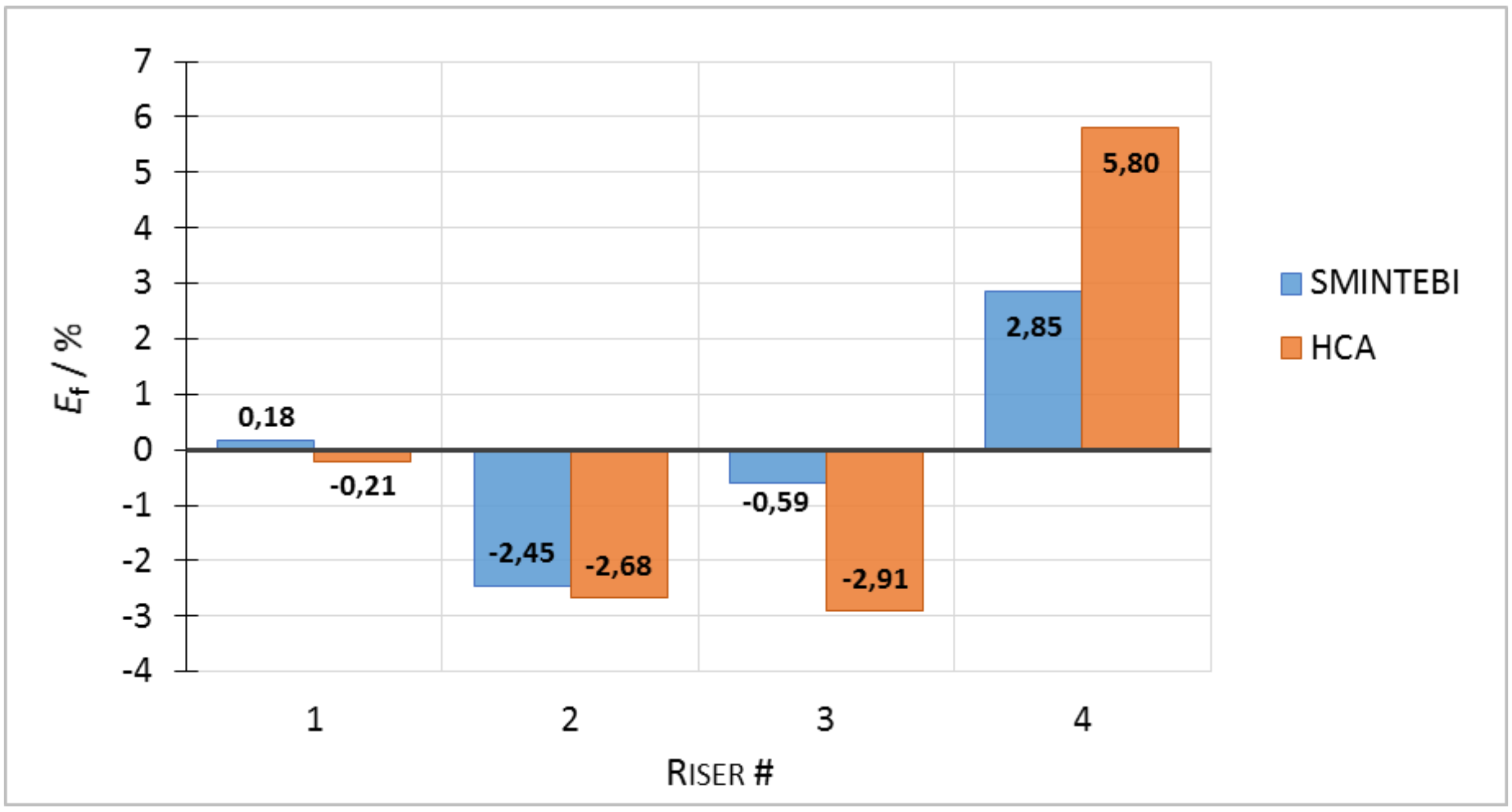

**Fig. 11.** SMINTEBI vs HCA heat allocation errors for the real operational site test.

**Table 11.** SMINTEBI vs HCA global indicators of performance for the real operation test.

| Global indicators | Real operational test | |
|---|---|---|
| | HCA | SMINTEBI |
| $\sigma(E_f)$ / % | 4.06 | 2.20 |
| $max(E_f)$ / % | 5.80 | 2.85 |
| $min(E_f)$ / % | -2.91 | -2.45 |
| $MAPE$ / % | 12.53 | 6.66 |
| $P_L$ / % | - | 100 |
| $\Delta E_{HCA}$ / % | - | -5.53 |

## 5. Conclusions

The proposed novel heat accounting method is able to provide the online calibration of the characteristic thermal model parameters of heating bodies installed in apartment buildings, by the analysis of the data of a smart sensors network (composed of smart thermostatic valves or heat cost allocators for instance). Such a capability to adapt radiator parameters to test site conditions represents the novelty of the proposed method, which significantly reduces the negative impact on the accuracy of conventional indirect heat accounting systems due to the deviation of actual installation and heat transfer conditions of heating bodies with respect to the nominal ones (e.g. reference installation and heat transfer conditions defined by the EN 442 Standard). In addition, the method can avoid the errors on heat allocation due to possible malfunctions of heating bodies (e.g. air entrainment and mud deposition inside the radiators), and wrong installation by adapting the corresponding thermal model parameters even to these particular situations.

The approach is data driven and based on the identification of the thermal model parameters of heating bodies, by solving the energy balance of the thermal-hydraulic distribution as a linear inverse problem, by means of a Tikhonov regularization (Regularized Least Squares) algorithm. The algorithm's inputs are the measurement data coming from the indirect heat accounting devices and the direct heat meter installed on the main supply line of the thermal hydraulic distribution, which are acquired regularly, e.g. daily, over the heating season.

Tests carried out at INRIM, where the method was validated by comparison against heat accounting results provided by reference direct heat meters, have demonstrated its capability to identify the proper thermal model parameters of radiators according to their actual installation. Specifically, as observed for the application of the method to a typical HCA system, the relative deviations between the estimated thermal model parameters and their nominal values (i.e. the values applied by conventional methods) range from 10% to 40%. This improvement produces a reduction of the error on the individual heating bill by up to about 50%, compared to that obtained by conventional HCA systems. Similar results have been obtained in the tests performed in a real building. In addition, tests performed on the SMINTEBI system have pointed out the effectiveness of the method even when applied to a heat accounting system that does not require any temperature measurement of radiator surface, but only the measurements provided by smart radiator thermostatic valves, i.e. the measurements of radiator inlet temperature and air temperature in proximity of the radiator valve. The main advantage of the SMINTEBI system is to provide heat accounting and temperature control at once by the same device.

The results of the experimental validation have demonstrated the effectiveness of the innovative heat accounting method in establishing a smart sensor network among the indirect heat accounting devices installed in centralized heating systems, outperforming conventional indirect heat accounting systems.

## Acknowledgements

This research was co-financed by Innosuisse – Swiss Innovation Agency (project 33395.1 IP-EE: SMINTEBI-MVP) & BKW.

## Appendix A. Uncertainty analysis

In order to evaluate the uncertainty associated to the estimations of the share of thermal energy consumption provided by the heat accounting systems, the different uncertainty contributions are evaluated and combined, accordingly with the propagation of uncertainties [24], by the measurement models described in Section 3.3.

### *A.1. Uncertainty of reference thermal energy measurements provided by radiator DHMs*

Considering the measurement model of equation (11), and assuming flow rate and temperature measurements as uncorrelated, the standard uncertainty associated to the reference thermal energy measurement $u(Q_{ref})$ provided by radiator DHMs can be evaluated as:

$$u(Q_{ref}) = Q_{ref}\sqrt{\left[\frac{u(\dot{V})}{\dot{V}}\right]^2 + \left[\frac{u(\Delta T)}{\Delta T}\right]^2 + \left[\frac{u(\rho)}{\rho}\right]^2 + \left[\frac{u(c_p)}{c_p}\right]^2 + \left[\frac{u_{co}}{Q_{ref}}\right]^2} \quad \text{(A. 1)}$$

where the standard uncertainty contribution $u_{co}$, associated to the thermal energy correction for flow rates lower than the flow meter cut-off, is calculated considering the thermal power correction as uniformly distributed between zero and $2{\cdot}q_{co}$:

$$u_{co} = \frac{\int q_{co}\, dt}{\sqrt{3}} \tag{A. 2}$$

The relative uncertainty contributions $u_{co}/Q_{ref}$ obtained in the test were lower than 1% for radiator heat consumptions higher than 90 kWh, that is for the 92% of radiators; for the other radiators, characterized by a heat consumption between 60 kWh and 70 kWh, the relative uncertainty contribution due to the low flow correction ranges from 2% to 6.5%.

The relative standard uncertainty associated to radiator flow rate measurements at water temperatures from (30 to 80) °C is equal to 0.15% for flow rates higher than 40 L $h^{-1}$ and increases up to 1% for lower flow rates (down to the low flow cut-off limit). The standard uncertainty associated to radiator inlet-outlet temperature difference measurements is equal to 0.04 °C. The relative uncertainties associated to the density and specific heat capacity of water are evaluated taking into account the uncertainty contributions associated to water temperature and pressure measurements, the uncertainty of the equation of state used for the determination of the thermodynamic properties of water [23], and the uncertainty associated to the measurements of density and specific heat capacity of the water used in the heating system at ambient temperature and pressure. The relative standard uncertainties associated to density and specific heat capacity of water have been evaluated equal to 0.05% and 0.75% respectively. Finally, the uncertainty contribution associated to the time integration of radiator thermal power can be considered negligible with respect to the other contributions, because of the low sampling time (15 s) used as integration time step for thermal energy measurement, which allows detecting the typical flow and temperature transients caused by STVs operation, with the appropriate time resolution.

### *A.2. Uncertainty of allocation units totalised by HCAs*

The uncertainty associated to the number of allocation units totalised by HCAs, $u(R)$, is evaluated considering i) the resolution of the counter, and ii) the relative uncertainty contribution $u_D$ given by the standard deviation of a rectangular distribution limited by the maximum relative display deviations of HCAs. The latter are established by the European Standard EN 834 for specific operating conditions, depending on the temperature difference $\Delta T_S$ between the surface of the radiator and the surrounding environment; for instance, the maximum relative display deviation is ±5% in the range of $\Delta T_S$ between 15 K and 40 K [8]. Therefore,

$$u(R) = R\sqrt{2\left(\frac{1}{R\,2\sqrt{3}}\right)^2 + u_D^2} \tag{A. 3}$$

where $R$ is the number (integer) of heat allocation units totalised by the HCA. The relative uncertainty contribution due to the resolution of HCAs was lower than 0.5% for about 92% of radiators, whereas the remaining 8% of radiators was characterized by a relative uncertainty contribution due to the HCA resolution lower than 1.5%. It is worth observing that the nominal thermal power of radiators, evaluated according to the EN 442 Standard and used to configure the HCAs, is characterized by a relative uncertainty that can be considered as already included in $u_D$, since the maximum permissible errors established by the EN 834 Standard for HCA readings take into account the uncertainty of HCA configuration parameters.

### *A.3. Uncertainty of radiator heat consumptions estimated by the improved heat accounting method*

The uncertainty associated to the estimations of radiator heat consumption obtained by the improved heat accounting method is evaluated considering i) the relative uncertainty contribution $u_H$ associated to the measurement of the overall thermal energy exchanged by the heat transfer fluid, ii) the relative uncertainty contribution $u_K$ associated to the prior knowledge $\boldsymbol{\theta_0}$ of radiator parameters (considered equal to the relative uncertainty associated to the nominal thermal power of radiators according to the EN 442 Standard), and iii) the relative uncertainty contribution $u_P$ associated to the estimates $\hat{\theta}_j$ of radiator parameters. The influence of the uncertainty sources related to heat losses through the pipework and transient heat transfer conditions is considered as an additional contribution to the global uncertainty of the heat accounting system, as described in Section 4.1.1.1. Thus, the standard uncertainty associated to the estimations of radiator heat consumption $u(Q_{rad})$ obtained by the improved heat accounting method can be evaluated as follows:

$$u(Q_{rad}) = Q_{rad}\sqrt{u_H^2 + u_K^2 + u_P^2} \qquad \text{(A. 4)}$$

The relative uncertainty $u_K$ is evaluated as the standard deviation of a rectangular distribution limited by the maximum permissible residuals of the EN 442 model for the evaluation of radiator thermal power; in particular, according to the EN 442 Standard, the residuals of the model shall be within ±2% of the measured radiator thermal output [11]. The relative uncertainty contribution $u_H$ associated to the direct measurement of the overall thermal energy exchanged by the heat transfer fluid is calculated combining the uncertainties of flow meter, temperature sensors pair, and calculation unit sub-assemblies, and was equal to about 1%, considering the overall flow and temperature measurement conditions during tests. The relative uncertainty contribution $u_P$ associated to the estimates of radiator thermal model parameters is evaluated for each radiator as the square root of the corresponding diagonal element of the covariance matrix $\boldsymbol{C_{\hat{\theta}}}$ given by equation (8), divided by the square root of the number of energy samplings and normalized with respect to the estimate $\hat{\theta}_j$ of the radiator parameter. The relative uncertainty $u_P$ also includes the contribution associated to the resolution of indirect heat accounting devices and ranges from 0.3% to 3% depending on the number of energy samplings considered in the batch of data for parameters estimation, and on radiators heat consumption.

### *A.4. Uncertainty of fractions of heat consumption and heat allocation errors*

The standard uncertainty associated to the fractions of heat consumption $u(f_{X,s})$ can be evaluated as:

$$u(f_{X,s}) = f_{X,s}\sqrt{\left[\frac{u(X_s)}{X_s}\right]^2 + \left[\frac{u\left(\sum_{s=1}^{N_s} X_s\right)}{\sum_{s=1}^{N_s} X_s}\right]^2 - 2\,\frac{cov\left(X_s\,, \sum_{s=1}^{N_s} X_s\right)}{X_s \sum_{s=1}^{N_s} X_s}} \qquad \text{(A. 5)}$$

where the standard uncertainty associated to the overall sum of heat consumptions $u\left(\sum_{s=1}^{N_s} X_s\right)$ has been evaluated assuming a null correlation between measurements or estimations of individual thermal energy consumptions associated to different subsets of radiators:

$$u\left(\sum_{s=1}^{N_s} X_s\right) = \sqrt{\sum_{s=1}^{N_s} u^2(X_s)} \quad \text{(A. 6)}$$

Similarly, for each *s*-th subset of radiators:

$$u(X_s) = \sqrt{\sum_{j=1}^{N} u^2\left(X_{j,s}\right)} \quad \text{(A. 7)}$$

Under the same assumption of uncorrelated thermal energy measurements or estimations, the covariance between individual and overall heat consumption has been evaluated as:

$$cov\left(X_s\,, \sum_{s=1}^{N_s} X_s\right) = u^2(X_s) \quad \text{(A. 8)}$$

Finally, the standard uncertainty associated to heat allocation errors can be evaluated as:

$$u\left(E_{f,X,s}\right) = \sqrt{u^2\left(f_{ref,s}\right) + u^2\left(f_{X,s}\right)} \quad \text{(A. 9)}$$

## References

[1]. Buildings - European Commission, available online: https://ec.europa.eu/energy/topics/energy-efficiency/energy-efficient-buildings_en

[2]. Energy Efficiency Directive (EED) European Commission, Directive 2012/27/EU of the European Parliament and of the Council of 25 October 2012 on energy efficiency, amending Directives 2009/125/EC and 2010/30/EU and repealing Directives 2004/8/EC and 2006/32/EC, 2012.

[3]. Directive (EU) 2018/844 of the European Parliament and of the Council of 30 May 2018 amending Directive 2010/31/EU on the energy performance of buildings and Directive 2012/27/EU on energy efficiency, 2018.

[4]. OIML R 75-1, Heat Meters, Part 1: General Requirements, International Organization of Legal Metrology, Paris, 2002.

[5]. EN 1434-1, Heat Meters, Part 1: General Requirements, European Committee for Standardization, Brussels, 2007.

[6]. MID directive European Commission, Directive 2014/32/EU of the European parliament and of the council of 26 February 2014 on the harmonisation of the laws of the Member States relating to the making available on the market of measuring instruments (recast), 2014.

[7]. L. Canale, M. Dell'Isola, G. Ficco, T. Cholewa, S. Siggelsten, I. Balen, A comprehensive review on heat accounting and cost allocation in residential buildings in EU, Energy and Buildings 202 (2019).

[8]. EN 834, Heat Cost Allocators for the Determination of the Consumption of Room Heating Radiators. Appliances with Electrical Energy Supply, European Committee for Standardization, Brussels, Belgium, 2013.

[9]. Italian National Unification UNI 11388:2015, Heat costs allocation system based on insertion time compensated by average temperature of the fluid (in Italian language only), 2015.

[10]. EN 442-1 Radiators and Convectors, Part 1: Technical Specification and Requirements, European Committee for Standardization, Brussels, Belgium, 2014.

[11]. EN 442-2 Radiators and Convectors, Part 2: Test Methods and Rating, European Committee for Standardization, Brussels, Belgium, 2014.

[12]. R. Marchesi, F. Rinaldi, C. Tarini, F. Arpino, G. Cortellessa, M. Dell'Isola, G. Ficco, Experimental analysis of radiators' thermal output for heat accounting, Thermal Science 23(2PartB) (2019) 989-1002.

[13]. F. Arpino, G. Cortellessa, M. Dell'Isola, G. Ficco, R. Marchesi, C. Tarini, Influence of installation conditions on heating bodies thermal output: preliminary experimental results, Energy Procedia 101 (2016) 74-80.

[14]. F. Saba, V. Fernicola, M.C. Masoero, S. Abramo, Experimental Analysis of a Heat Cost Allocation Method for Apartment Buildings, Buildings 7 (2017) 20.

[15]. Italian National Unification UNI 10200:2018, Impianti termici centralizzati di climatizzazione invernale e produzione di acqua calda sanitaria – Criteri di ripartizione delle spese di climatizzazione invernale ed acqua calda sanitaria (in Italian language only), 2018.

[16]. S. Nižetić, N. Djilali, A. Papadopoulos, J. J.P.C. Rodrigues, Smart technologies for promotion of energy efficiency, utilization of sustainable resources and waste management, Journal of Cleaner Production 231 (2019) 565-591.

[17]. A.N. Tikhonov, A. Goncharsky, V.V. Stepanov, A.G. Yagola, Numerical methods for the solution of ill-posed problems, Springer Netherlands, 1995.

[18]. L. Celenza, M. Dell'Isola, G. Ficco, B.I. Palella, G. Riccio, Heat accounting in historical buildings, Energy and Buildings 95 (2015) 47–56.

[19]. G. Ficco, L. Celenza, M. Dell'Isola, P. Vigo, Experimental comparison of residential heat accounting systems at critical conditions, Energy and Buildings 130 (2016) 477-487.

[20]. P.C. Hansen, Analysis of Discrete Ill-Posed Problems by Means of the L-Curve, SIAM Review 34 (4) (1992) 561-580.

[21]. P.C. Hansen, The L-curve and its use in the numerical treatment of inverse problems, In Computational Inverse Problems in Electrocardiology, ed. P. Johnston, Advances in Computational Bioengineering, WIT Press (2000) 119-142.

[22]. IAPWS 2007 Revised Release on the IAPWS Industrial Formulation 1997 for the Thermodynamic Properties of Water and Steam, International Association for the Properties of Water and Steam, Lucerne, Switzerland, 2007.

[23]. IAPWS Advisory Note No. 1 Uncertainties in Enthalpy for the IAPWS Formulation 1995 for the Thermodynamic Properties of Ordinary Water Substance for General and Scientific Use (IAPWS-95) and the IAPWS Industrial Formulation 1997 for the Thermodynamic Properties of Water and Steam (IAPWS-IF97), International Association for the Properties of Water and Steam, Vejle, Denmark, 2003.

[24]. BIPM, IEC, IFCC, ISO, IUPAC, IUPAP and OIML 2008 Evaluation of Measurement Data - Guide to the Expression of Uncertainty in Measurement, Joint Committee for Guides in Metrology, Bureau International des Poids et Mesures, JCGM 100.